\documentclass[useAMS,usenatbib,referee]{mn2e}

\usepackage{lscape}
\usepackage{graphicx}
\usepackage{txfonts}
\usepackage{longtable}

\title[SuperWASP Observations of 17P/Holmes]{SuperWASP Observations of the 2007 Outburst of Comet 17P/Holmes}

\author[H. H. Hsieh, A. Fitzsimmons, Y. C. Joshi, D. J. Christian, D. L. Pollacco]
       {Henry H. Hsieh$^1$\thanks{Email: h.hsieh@qub.ac.uk}, Alan Fitzsimmons$^1$, Yogesh Joshi$^{1,2}$, Damian Christian$^{1,3}$
        and Don L. Pollacco$^1$\\
        $^1$ Astrophysics Research Centre, Queen's University, Belfast, BT7 1NN, United Kingdom\\
        $^2$ Aryabhatta Research Institute of Observational Sciences, Manora Peak, Nainital 263129, India\\
        $^3$ Physics \& Astronomy Department, California State University Northridge, Northridge, California 91330-8268, USA
         }

\begin{document}

\date{Submitted, 2010 Feb 3; Accepted, 2010 May 12}

\pagerange{\pageref{firstpage}--\pageref{lastpage}} \pubyear{2010}

\maketitle

\label{firstpage}

\begin{abstract}
We present wide-field imaging of the 2007 outburst of Comet 17P/Holmes
obtained serendipitously by SuperWASP-North on 17 nights over a
42-night period beginning on the night (2007 October 22-23)
immediately prior to the outburst.  Photometry of 17P's unresolved coma
in SuperWASP data taken on the first night of the outburst is consistent
with exponential brightening, suggesting that the rapid increase in the
scattering cross-section of the coma could be largely due to the progressive
fragmentation of ejected material produced on a very short timescale at
the time of the initial outburst, with fragmentation timescales decreasing
from $t_{frag}\sim2\times10^3$~s to $t_{frag}\sim1\times10^3$~s over our
observing period.  Analysis of the expansion of 17P's coma reveals a velocity
gradient suggesting that the outer coma was dominated by material ejected
in an instantaneous, explosive manner.  We find an expansion
velocity at the edge of the dust coma of $v_{exp}=0.55\pm0.02$~km~s$^{-1}$ and
a likely outburst date of $t_0=2007~{\rm October}~23.3\pm0.3$, consistent with
our finding that the comet remained below SuperWASP's detection limit
of $m_V\sim15$~mag until at least 2007 October 23.3.  Modelling of 17P's gas coma
indicates that its outer edge, which was observed to extend past the outer
dust coma, is best explained with a single pulse of gas production,
consistent with our conclusions concerning the production of the outer dust coma.
\end{abstract}


\section{INTRODUCTION}

Discovered on 1892 November 9 by Edwin Holmes, Comet 17P/Holmes (hereafter,
17P) is a dynamically ordinary Jupiter-family comet, with a semimajor
axis of $a=3.617$~AU, an eccentricity of $e=0.432$, an inclination of
$i=19.11^{\circ}$, an orbital period of $P_{orb}=6.8$~years, and a
Tisserand parameter (with respect to Jupiter) of $T_J=2.859$.  On 2007
October 24.1, it was discovered by J. A. Henr\'iquez Santana to be undergoing
a substantial outburst.  Starting from an initially observed magnitude of
$m_V\sim8.4$ \citep[compared to a pre-outburst brightness of $m_V\sim17$ measured
on 2007 October 23.1;][]{cas07}, it brightened by approximately
0.5 mag~hr$^{-1}$ over the course of 6 hours \citep{buz07}.  The comet reached
$m_V\sim2.0$ by 2007 October 25.1 \citep{spo07,elh10}, representing a million-fold
increase in brightness over only 2 days.  This outburst of 17P mirrored
a similar outburst that led to the comet's initial discovery in 1892, and
continued to be monitored by both amateur and professional astronomers as
it progressed.

Spectroscopic analyses indicated that the coma was largely dominated by dust
grains \citep{sch07}, but evidence of cometary volatiles and sublimation
byproducts, including OH, H$_2$O, NH, CN, HCN, HNC, C$_2$, C$_3$, C$_2$H$_2$, C$_2$H$_6$,
CS, CH$_3$CN, and CH$_3$OH, were also detected during
the outburst \citep{dra07,fit07,kob07,sch07,wag07,boc08,del08,yan09}.
Negative polarisation was found for the dust coma by \citet{jos10}, consistent with
measurements made for other comets.
A delay in the appearance of CN, C$_2$, and [OI] emission lines between
2007 October 24.58 and October 25.46 led
\citet{kob07} to speculate that their observations could be at least partially
explained by the delayed sublimation of icy grains ejected at the time of the
outburst.  This hypothesis is supported by \citet{del08} who found evidence of a
distributed source of sublimation products, consistent with
the ejection and subsequent delayed sublimation of icy grains in the coma.
The total mass loss due to this outburst has been estimated to be $\sim10^{12}$ kg
or a few percent of the comet's total mass \citep{sek08a,mon08}.

\section{OBSERVATIONS AND REDUCTION}

The Wide Angle Search for Planets (WASP) operates two facilities:
SuperWASP-N, located in the northern hemisphere on La Palma, and
WASP-S, located in the southern hemisphere at the South African
Astronomical Observatory.  These two facilities, comprising eight
cameras each, provide fully robotic, high time-resolution, and ultra-wide
angle imaging of selected star fields for the primary purpose of detecting
transiting extrasolar planets.  Each camera employs 200mm f/1.8 Canon lenses
with apertures of 11.1~cm and a broadband filter defining a custom passband from
400 to 700~nm. The detectors are $2048\times2048$ thinned e2v CCDs with an image
scale of $13\farcs7$~pixel$^{-1}$ for a total field of view of $7.8^{\circ}$
by $7.8^{\circ}$ per camera.  Images are processed by an automated reduction pipeline
that performs bias and thermal dark frame subtraction and flatfield
reduction \citep{pol06}.

\begin{figure}
\includegraphics[width=5.5in]{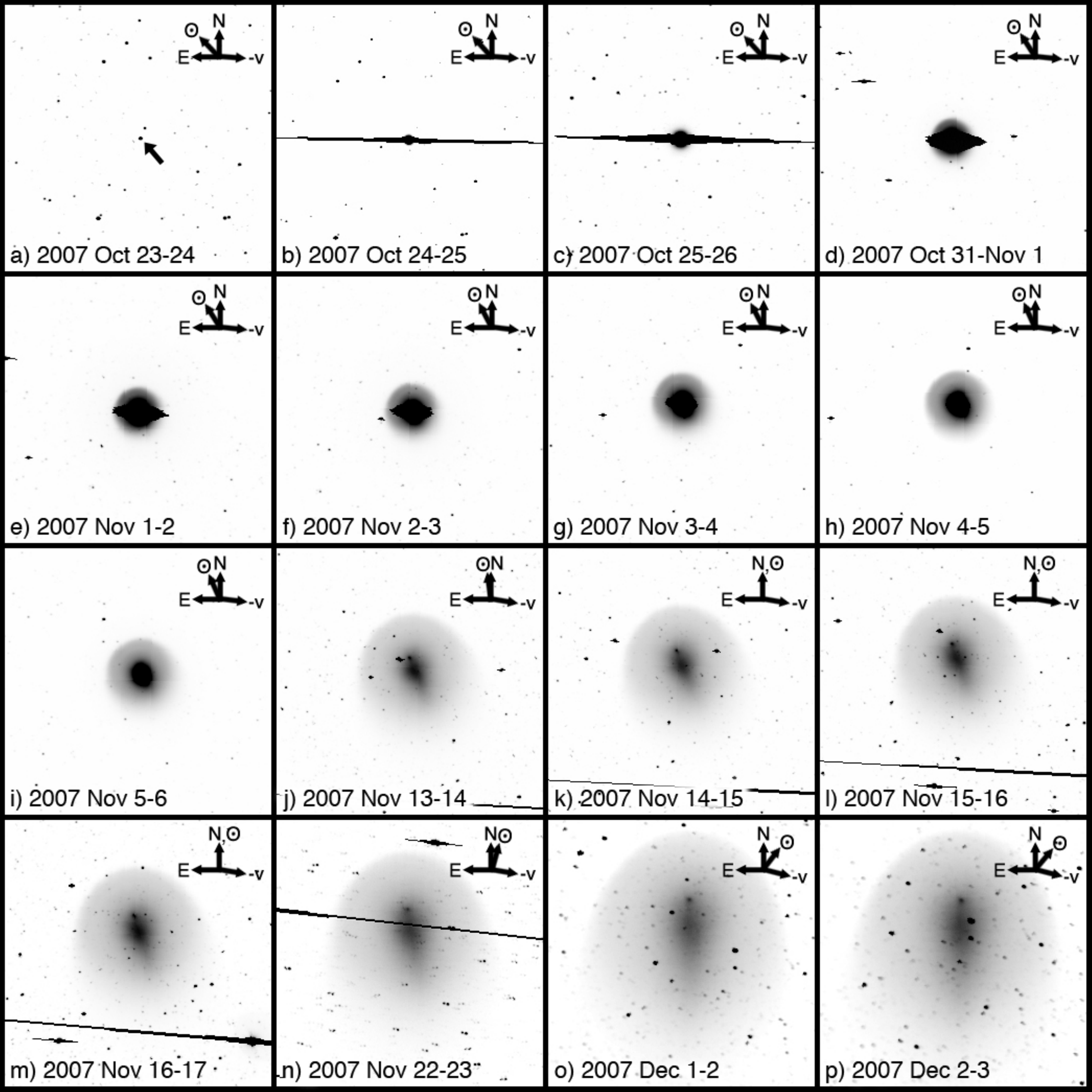}
\caption{\small Images of 17P on 16 nights between 2007 Oct 23
  and 2007 Dec 3 using SuperWASP-N.  Each image comprises 30~s of exposure
  time and is 1$^\circ$ by 1$^\circ$ in
  size.  Arrows indicate north (N), east (E), the direction towards the
  Sun ($\odot$), and the negative heliocentric velocity vector (-v).
  Solid lines intersecting various images are saturated CCD columns from the
  comet itself or nearby bright field stars.
  }
\label{images_all}
\end{figure}

SuperWASP-N (hereafter, SuperWASP) serendipitously observed 17P in one camera on
numerous nights during the comet's 2007 outburst, starting on the night
of 2007 October 23-24 and continuing through 2007 December 2-3, providing
us with a well-sampled, long-baseline data set for analysing the comet's
evolving brightness and morphology.  We list details of these observations in
Table~\ref{obslog} and show typical images of the comet obtained on each
night of observations in Figure~\ref{images_all}.  Data from the
night prior to the outburst (2007 October 22-23) is also available.
Inspection of these data shows no evidence of the comet at its expected
position, indicating that the comet remained below SuperWASP's detection
threshold ($m_V\sim15$~mag) until at least 2007 October 23.3.

On the first night of 17P's outburst, a total
of 84 images, comprising 30 seconds of exposure time each, were obtained
from 2007 Oct 23.9851 (UT 23:38) to Oct 24.2681 (UT 06:26).
Images were obtained in consecutive pairs
approximately every 10 to 12 minutes.  The steadily-brightening comet
was visible from the start of the observations but became saturated on
the CCD from approximately Oct 24.1496 onwards, leaving 44 images from which
reliable photometry could be obtained.  The FWHM measured for field star
surface brightness profiles during the period of observations was
approximately $23\farcs3$.
Aperture photometry was performed on the comet using apertures with radii of $70\farcs0$.
The background sky flux was measured at several nearby locations on the sky
where field star contamination was minimal.
Five nearby field stars (within 30 arcmin of the comet)
were also measured using the same apertures and used
to make differential photometric corrections to our measured comet fluxes
to ensure a consistent photometric baseline throughout the observation period.

Because SuperWASP employs a non-standard wide filter optimised for its
primary function as a transit detector \citep{pol06}, precise absolute
photometry of 17P in a standard passband is difficult to obtain
from this data set. 
We estimate apparent magnitudes by assuming the spectrum of 17P
to be approximately solar and performing differential photometry using nearby FGK-type field
stars whose magnitudes and spectral types are known (Table~\ref{fieldstars}). We assign a combined systematic and measurement uncertainty
of $\sim$0.2~mag to these measurements.
These magnitudes are computed for comparison to other published results only.
All analysis discussed hereafter is performed using differentially-calibrated instrumental fluxes (as described above)
unless otherwise specified, and as such is
unaffected by uncertainties in these magnitude calibration calculations.
All photometry results are shown in Table~\ref{photometry}.

\section{DATA ANALYSIS}

\subsection{Lightcurve\label{initlightcurve}}

\subsubsection{Coma Optical Depth}

If we assume that the coma of 17P is dominated by dust grains and is
optically thick on the first night of the comet's outburst (2007~October 23-24),
we would expect its brightness ($I$) to vary linearly
with the surface area of the coma as projected on the sky.  If the coma
radius increases at a linear rate, its surface area will
vary as the square of elapsed time, $t-t_0$,
since the outburst, giving the following expression for the brightness:
\begin{equation}
I = I_0 + z(t-t_0)^k
\label{squarelaw}
\end{equation}
where $k=2$ and $z$ is a scaling constant.
Assuming an effective initial brightness, $I_0\sim150$~ADU (equivalent to a
$\sim$0.05$\sigma$ signal in our data),
consistent with reported pre-outburst magnitudes of 17P of $m_V\sim17.0$
\citep{cas07} and fitting only to the first $\sim$2 hours of SuperWASP data,
we find best-fit values of $z=2.3\times10^6$  
and $t_0= 2007~{\rm October}~23.8\pm0.1$,     
as reported in \citet{hsi07}. This value for $t_0$ is consistent with
the outburst time initially estimated by \citet{gai07}.

While this early photometry
initially appears to be consistent with a linearly expanding, optically
thick coma, combining these data with our analysis of the coma's expansion
velocity (described below in \S\ref{comaexp}) shows that this is likely not the
case.  We can estimate the total scattering cross-section, $A=\pi r_e^2$,
of the particles in the coma from
\begin{equation}
p_Vr_e^2 = 2.24\times10^{22}\times10^{0.4[m_{\odot}-m(1,1,0)]}
\end{equation}
where $p_V$ is the geometric $V$-band albedo, assumed here to be $p_V=0.04$,
and $m_{\odot}=-26.66$ is used for the absolute solar $V$-band magnitude.
Assuming a phase coefficient of $\beta = 0.035$~mag~deg$^{-1}$, we obtain
starting and ending absolute magnitudes of $m_{start}(1,1,0)=6.08$ and
$m_{end}(1,1,0)=5.50$ for the period of observations for which the
comet's brightening appears to follow a square law,
giving starting and ending total scattering
cross-sections of $A_{start}=1.41\times10^5$~km$^2$ and
$A_{end}=2.41\times10^5$~km$^2$.  If the coma is optically thick, meaning
that the observed scattering cross-section is equal to the total area of the
coma, $A_{coma}$, we obtain starting and ending coma
radii of $r_{start}=212$~km and $r_{end}=277$~km, assuming a
circular coma.  Our first and last photometry measurements are separated
by 6487~seconds, so the assumption of an optically thick coma
yields a coma expansion velocity of only 0.01~km~s$^{-1}$.

This velocity is far slower than our calculated coma expansion velocity and
all other reported expansion velocities ({\it cf.}
\S\ref{comaexp}).  Using a smaller assumed albedo value for the coma
dust particles has the effect of increasing the total effective scattering
cross-section, therefore increasing the implied expansion velocity.
An implausibly low assumed albedo of $p_V=1.6\times10^{-5}$ is required,
however, to produce an expansion velocity of 0.5~km~s$^{-1}$ (where
$r_{start}=1.06\times10^4$~km and $r_{end}=1.38\times10^4$~km).
Assuming a larger phase coefficient value also increases the implied
expansion velocity by increasing the implied brightness of the coma at
$\alpha=0^{\circ}$, but again, an implausible value for the phase
coefficient ($\beta=0.53$~mag~deg$^{-1}$) is required to produce an
expansion velocity of 0.5~km~s$^{-1}$, assuming
$p_V=0.04$.  In fact, as shown in Figure~\ref{betaalbedo}, given the range
of cometary phase coefficients \citep[$0.01<\beta<0.09$;][]{mee87,sno08}
and albedos \citep[$0.01 < p_V < 0.07$;][]{lam04} measured to date,
no plausible combination of $\beta$ and $p_V$ is able to account for our
photometry data given the assumption of an optically-thick coma with an outer
edge expanding at 0.5~km~s$^{-1}$.

The implausible conditions necessary for 17P's coma to be optically
thick strongly suggest that the coma is optically thin even
in our earliest observations of the outburst, confirming the results
of \citet{sek09a}. This conclusion
is supported by our coma expansion analysis (\S\ref{comaexp}) which
suggests that the radius of the outer edge of the coma should have already
been $5.0\times10^4$~km at the time of SuperWASP's first image of the comet
on the first night of the outburst, making it approximately 200 times larger
than an optically-thick coma would be (assuming typical cometary albedo and
phase coefficient values of $p_V=0.04$ and $\beta=0.035$~mag~deg$^{-1}$).
We therefore find that the rapid brightening we observed for the comet on
2007 October 23-24 must have been caused by a process other than the expansion
of an optically thick coma.

\subsubsection{Evidence for Fragmentation}

If we assume that 17P's coma was already optically thin by the time of our
first observation, it is possible that most of the observed material
could have been emitted on a very short timescale around the time of the outburst,
and that the brightening of the coma was primarily due to the
fragmentation of the dust grains into smaller particles leading to
an increase in reflecting surface area.  This scenario was also suggested by \citet{gai07} and
\citet{kob07}.
Fragmentation of dust particles has also
been inferred from in-situ measurements at other comets such as 1P/Halley \citep{ho07}
and 81P/Wild 2 \citep{tuz04}, as well as many ground-based observations
of comets such as C/Hale-Bopp \citep{pit97}.

If a dust grain breaks into two daughter grains when it fragments, the original visible
cross-sectional area of the dust increases by a factor $f$.
As long as the fragmentation timescale, $t_{frag}$, and the albedo
are roughly size-invariant ({\it i.e.} they do not change as fragmentation progresses),
the dust's total visible cross-section, and 
hence its brightness, will
increase exponentially with the form
\begin{equation}
I(t-t_0)=I_0 e^{z(t-t_0)}\equiv I_0 f^{(t/t_{frag})}
\end{equation}
where $I_0$ is the brightness of the dust at an initial time $t_0$.
The dust grain fragmentation timescale $t_{frag}$ is related by the brightness exponent $z$ by
\begin{equation}
t_{frag}=\frac{\ln{f}}{z}
\end{equation}
For splitting into two daughter fragments, the maximum value of $f$ is
when the daughter grains have equal mass, giving $f(max)=1.26$.
For splitting into more than two daughter fragments, $t_{frag}$ will
be larger by a factor $1.44\ln N$, where $N$ is the number of simultaneously
produced daughter particles and we assume that all daughter particles have
equal mass.

\begin{figure}
\includegraphics[width=5.5in]{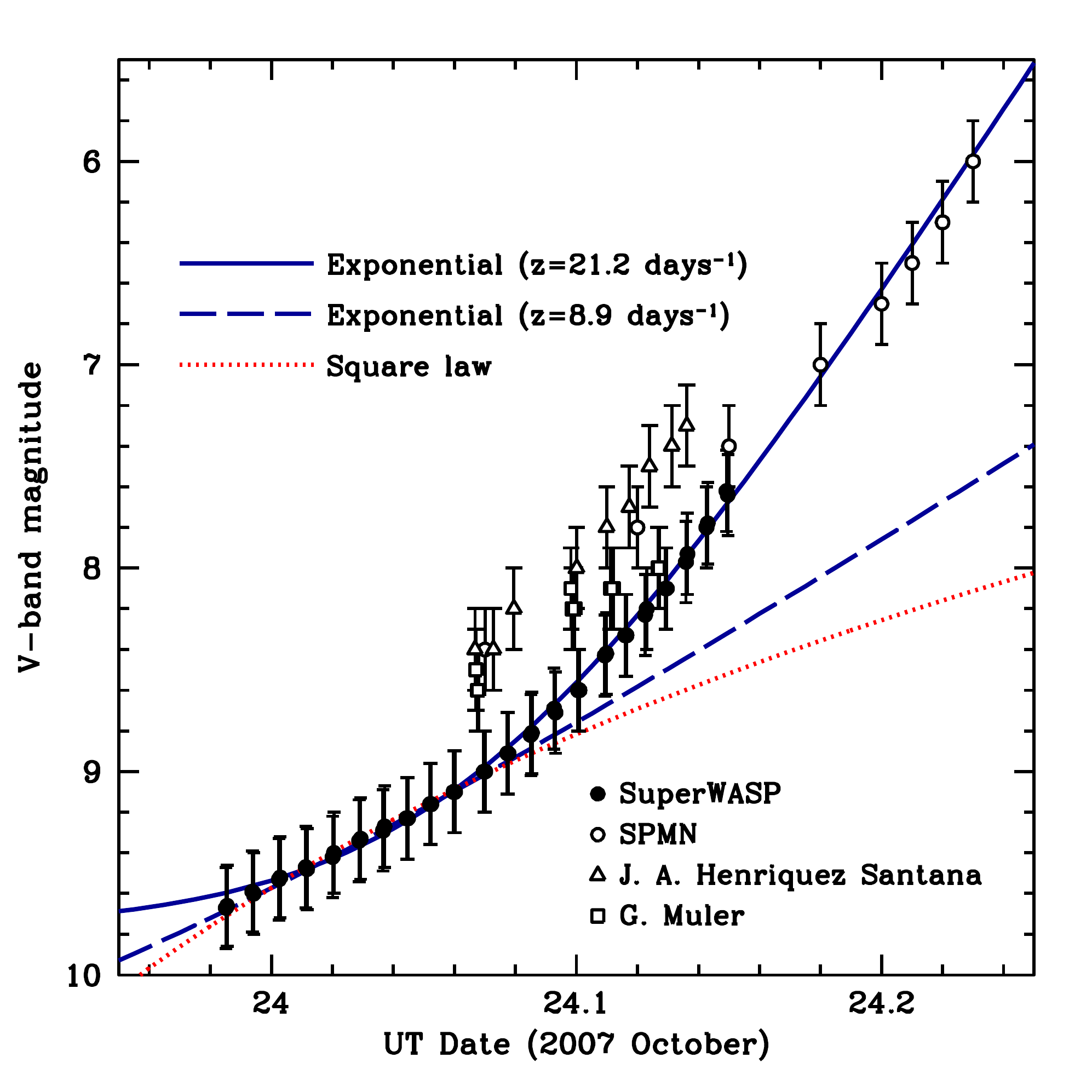}
\caption{\small $V$-band magnitudes (solid circles) measured for the nucleus of
  17P with the best-fit square-law function
  and best-fit exponential function plotted as a
  dashed line and a solid line, respectively.
  }
\label{fluxplot}
\end{figure}

\begin{figure}
\includegraphics[width=5.5in]{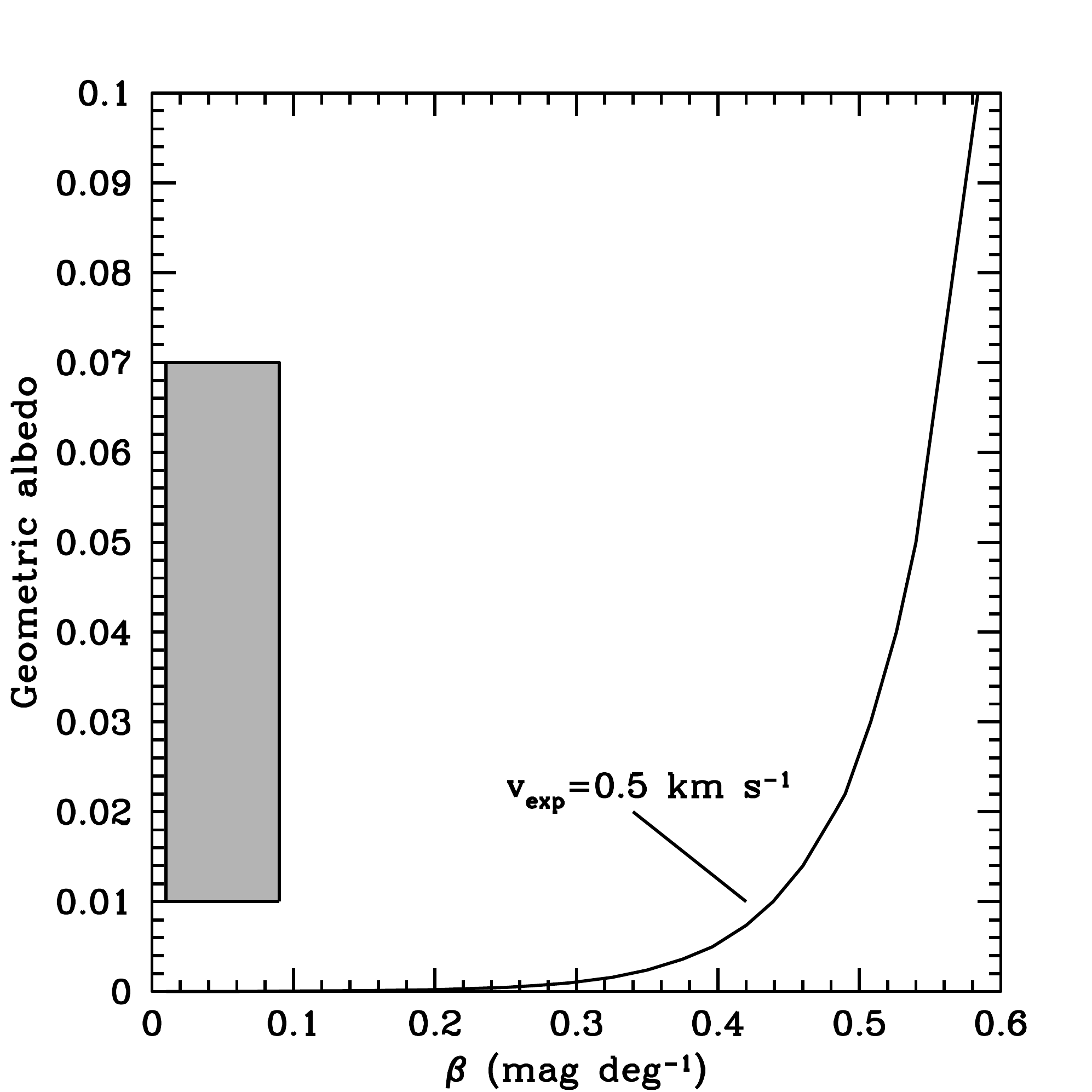}
\caption{\small The locus of values (plotted as a solid line)
  of $\beta$ and albedo
  required for the particles in an optically-thick coma with an
  outer edge expanding at 0.5~km~s$^{-1}$ in order to account for the
  observed lightcurve in Figure~\ref{fluxplot}.
  The shaded area
  indicates the range of cometary $\beta$ and albedo values measured to date.
  }
\label{betaalbedo}
\end{figure}

Analysing our data in the context of this fragmentation model,
we fit 17P's early lightcurve to the following function:
\begin{equation}
I = A (e^{z(t-t_0)}-1) + I_0
\label{expfunction}
\end{equation}
Mathematically, $A$ and $t_0$ in this function cannot be fit simultaneously,
meaning that Equation~\ref{expfunction} cannot be used to derive a unique outburst
time.  Thus, for this analysis, we assume an outburst time of $t_0=2007~{\rm October}~23.3$
as determined from the expansion of the coma (discussed below in \S\ref{comaexp}).
Fitting the entire $\sim$4 hours of unsaturated 17P data, we find best-fit parameters of
$A=(1.0\pm0.2)\times10^{-2}$~ADU,           
$z=21.2\pm0.2$~days$^{-1}$,                       
and $I_0=(1.10\pm0.02)\times10^5$~ADU     
(corresponding to $m_V\sim9.8$~mag).
An exponential function using these best-fit parameters (where fluxes have
been converted to equivalent $V$-band magnitudes), along with the square law
discussed above, is plotted in Figure~\ref{fluxplot}. 

Overlaying on Figure~\ref{fluxplot} other published photometry (Table~\ref{otherphotometry}) from
the Spanish Meteor and Fireball Network \citep[SPMN;][]{tri08}, and from J. A. Henr\'iquez
Santana and G. Muler \citep{spo07},
we find slight discrepancies between our computed
magnitudes and these other data. However given the unknown calibration methods of
Henr\'iquez Santana and Muler and the uncertain absolute calibration of both SuperWASP
and SPMN data (stemming from the use of non-standard filters at both facilities), 
 we judge our data and our derived best-fit exponential brightening function
to be approximately consistent with these other data.

This result implies that dust particles in 17P's coma had average fragmentation lifetimes of
$t_{frag}\lesssim10^3$ sec, assuming bilateral splitting. However, we note that
when fitting this exponential function to our data, 
the fit is noticeably poorer to the first half of the data set compared to
the second half.  Functionally, this discrepancy can be accounted for
if $z$ increases over our observing period.
In Fig.~\ref{fluxplot}, a second exponential function is also shown fitted only to
the first 2 hours of data with $z=8.9$~days$^{-1}$, $t_0=2007~{\rm October}~23.3$,
$A=200$~ADU, and $I_0=3.0\times10^4$~ADU corresponding to $m_V\sim11.2$~mag.
This gives a fragmentation timescale (assuming two daughter grains per
splitting event) of $t_{frag}\lesssim2\times10^3$~s.
The improved accuracy of such a two-exponential fit implies that as the
outburst progressed, the rate of fragmentation increased.
This behaviour could physically correspond to a situation in which the
outburst ejected clumps of dust bound together by ice.  As the ice sublimated
and became progressively unable to bind those clumps together,
increasingly rapid disintegration could have followed.

\subsection{Coma Expansion\label{comaexp}}

Using data taken from 2007 October 31
onward (when at least a portion of the coma had expanded beyond the central saturated core of the
comet in our data), we can measure the size of the dust coma on successive
nights and thus determine its expansion rate and, in principle, the
time of the outburst's onset.
 Two complications in this method as noted by
\citet{sek08a} are the presence of an extended gas coma with poorly-defined
edges that obscures the extent of the relatively more well-defined
dust coma ({\it cf}. \S\ref{gasmodel}), and the non-sphericity of the coma after approximately
2007 November 3.

Additionally, we note that as the coma expands and
fades, successive portions of the outer coma are rendered
unobservable as they approach the signal-to-noise detection
limit of the background sky.  This process complicates
measurements of the size and expansion rate of the coma since it means that
the visually-determined ``edge'' of the coma does
not correspond to the same parcels of coma dust at all times,
but instead simply represents the locus of points at
which the coma's surface brightness is detectable against the sky.
Even assuming the sky brightness to be approximately constant from
night to night (which it is not, due to the changing proximity and phase
of the Moon and the altitude of observation), since the surface brightness
of a given parcel of dust decreases as the coma expands and becomes
more diffuse, following a point of constant brightness will actually lead
steadily inward in the expanding reference frame of the coma.  Thus,
over time, simply measuring the coma's observable edge
will lead to increasing underestimates of the coma's true size,
leading to an underestimate of its true expansion velocity.

With these concerns in mind, we prepare our data for the measurement of
the coma size and expansion rate by rotating
comet images from each night to align the central
axis (as projected in the image plane) of the comet's elongated coma
with each image's vertical axis.  This procedure is a straightforward
matter for data obtained on 2007 November 13 onwards, in which the coma
is visibly elongated, and the position angle of the central axis is clear
(Table~\ref{obslog}). For data
obtained on 2007 November 6 or earlier, the coma is largely circular,
making it difficult to discern the orientation of its central
axis.  For these data, we simply adopt the antisolar vector (as projected in the
plane of the sky) as the coma's central axis, and apply image rotations accordingly.

\begin{figure}
\includegraphics[width=5.5in]{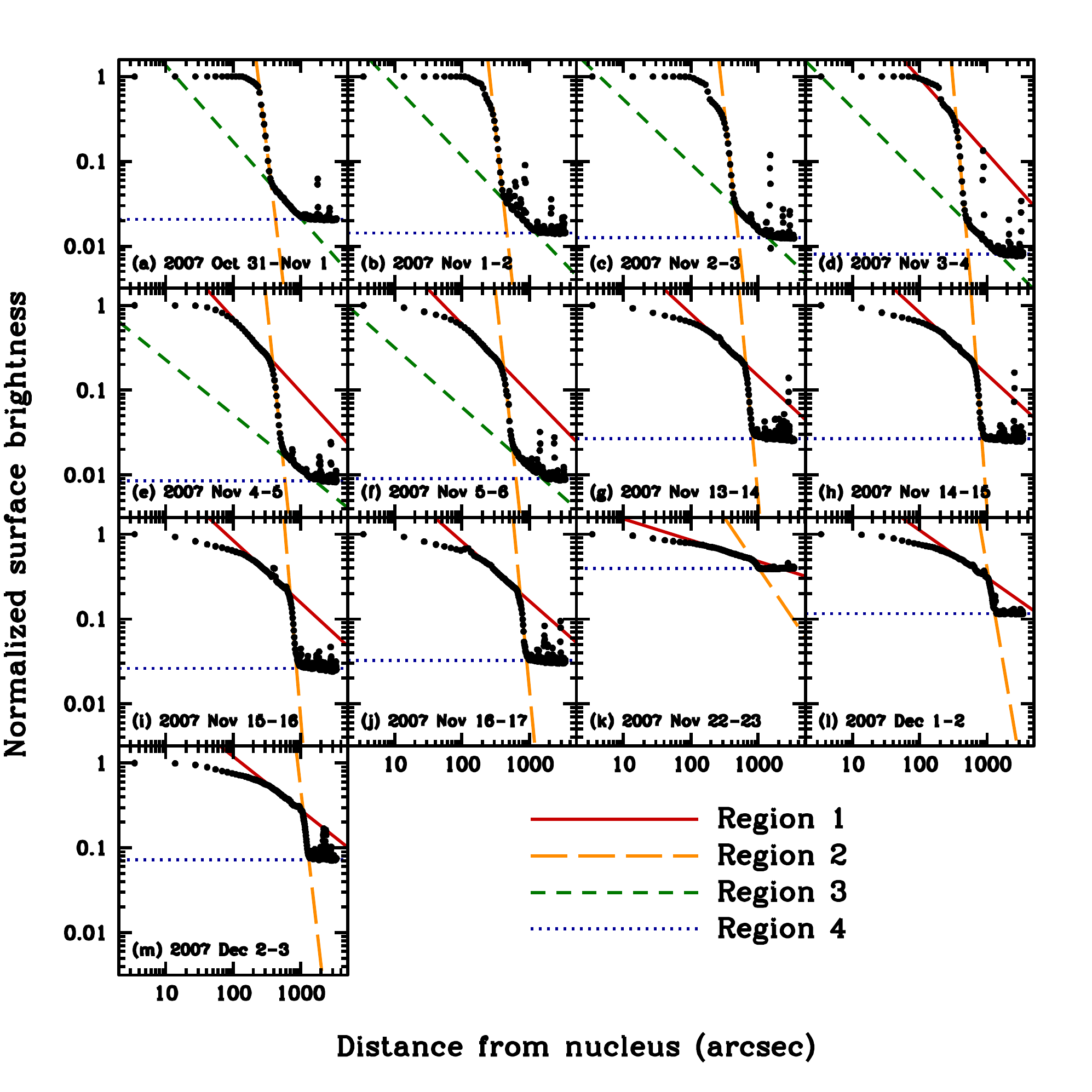}
\caption{\small Linear surface brightness profiles of the coma of
  17P as measured along a line passing through the nucleus and
  perpendicular to the coma's central axis, taken to be the
  antisolar vector (as projected in the sky)
  for (a)--(f) and the coma's axis of symmetry for (g)--(m).
  Measured normalised surface brightness values are plotted as solid dots,
  while linear fits to Regions 1, 2, 3, and 4 of each surface brightness
  profile (as described in the text) are plotted as solid lines,
  large-dashed lines, small-dashed lines, and dotted lines, respectively.
  Region 1 parameters are not calculated for Oct 31-Nov 3 due to extensive inner coma
  saturation on these dates.  Region 3 is absent for Nov 13-14 onwards.
  We note that with the exception of Nov 22-23 when the Moon was nearly full,
  the sky brightness is relatively constant (Table~\ref{lsbparams}.  The
  apparent increase in Region 4 over time is due to the decrease in peak coma
  brightness, to which each surface brightness profile is individually normalised.
  }
\label{lsbprofiles}
\end{figure}

For each rotated image, we measure the flux in the coma along a horizontal
three-pixel-wide strip passing through the comet nucleus, where the nucleus
is assumed to be at the centre of the nearly circular coma in data obtained
prior to 2007 November 5-6.
We then average the fluxes contained in pairs of pixels which are
equidistant from the nucleus on each side of that axis.
Finally, we normalise the results to the flux of the photometry aperture 
centred on the nucleus and plot these normalised fluxes as a function of
distance from the comet's nucleus for each image (Fig.~\ref{lsbprofiles}).
By constructing profiles in this way, we aim to minimise the
effects on our results of the coma's non-sphericity and
highly-projected orientation in the sky.

We note immediately that the surface brightness profiles we obtain
are of approximately the same form from night to night (with the exception of the
profile on 2007 November 22-23 whose anomalous appearance we attribute to
the close proximity of the nearly-full moon on that night).
This stability of our surface
brightness profiles gives us confidence that we are able to achieve
consistent results despite the large spatial scale of the coma and
widely-varying stellar backgrounds.  Significantly, we also note that
these linear surface brightness profiles do not match the $r^{-1}$ radial
surface brightness profile of a canonical spherically symmetric,
steady-state cometary dust coma.  Instead, profiles from
2007 October 31-November 1 through 2007 November 5-6
(Fig.~\ref{lsbprofiles}a - \ref{lsbprofiles}f) consist of saturated
pixels near the nucleus, followed by a region of moderate slope
($\sim r^{-0.6}$ to $r^{-0.9}$; hereafter ``Region 1'') leading into
an extremely steep section ($\sim r^{-5}$ to $r^{-11}$; hereafter
``Region 2''). Region 2 then transitions into another region of moderate
slope ($\sim r^{-0.7}$ to $r^{-0.9}$; hereafter ``Region 3'') which
finally becomes essentially horizontal ($\sim r^{0.0}$; hereafter
``Region 4'') far from the nucleus where the surface brightness profile
becomes dominated by the background sky. 
For data from
2007 November 13-14 onwards
(Fig.~\ref{lsbprofiles}g - \ref{lsbprofiles}m), the central portion of
the coma is no longer saturated, while Region 2 of each profile
transitions directly into Region 4 without passing through Region 3.
Where possible, we compute best-fit linear functions for each of these
regions in each profile (Fig.~\ref{lsbprofiles}, Table~\ref{lsbparams}).
Region 1 parameters are not calculated for 2007 October 31-November 3 due to extensive inner
coma saturation on these dates.

The observed evolution of 17P's coma profile is consistent with a
growing coma which maintains an approximately constant morphology
as it expands and fades.  We interpret Region 1 as the dust-dominated
portion of the coma, and Region 2 as the edge of the
dust coma, with its steep slope likely indicating a sharp cutoff at the
small end of the size distribution of visible grains.  The inconsistency
with the slope of Region 1 ($-0.9$, declining to $-0.6$)
with that of a canonical steady-state dust coma under radiation pressure ($-1.0$ to $-1.5$) indicates that
the coma is not in fact steady-state.
Such a steadily decreasing slope would in fact be expected if a single
outburst is assumed to account for the majority of the observed dust coma,
as faster, smaller dust grains in the outer coma will expand at a higher
velocity than slower-moving, large particles in the inner coma.
We hypothesise that the transient
Region 3 could be due to gas extending beyond the outermost edge of
the dust that otherwise dominates the coma profile.  We discuss this
hypothesis at length below (\S\ref{gasmodel}).  The fluctuating length of
Region 2 after the disappearance of Region 3 confirms our
concern that the intersection of the outer coma and the
sky is an unreliable determinant of coma size.

With the above issues in mind, we suggest that a more precise measure
of the expansion rate of 17P's coma can be achieved by 
tracking the motion of reference points {\it within} the coma.
Specifically, we suggest that the relatively
well-defined transitions between the various surface brightness profile
regions described above could be useful reference points. We note that
uncertainty is not entirely eliminated by the use of this method as the
shape of the coma profile does evolve slightly over time ({\it cf.}
Table~\ref{lsbparams}) meaning that our chosen reference points likely shift
slightly within the expanding reference frame of the coma. Nonetheless, given the
fundamentally problematic nature described above of measuring the expansion of a
progressively fading coma using brightness-dependent reference points, we believe tracking
morphology-dependent reference points to be the most reliable method at our disposal.

\begin{figure}
\includegraphics[width=5.5in]{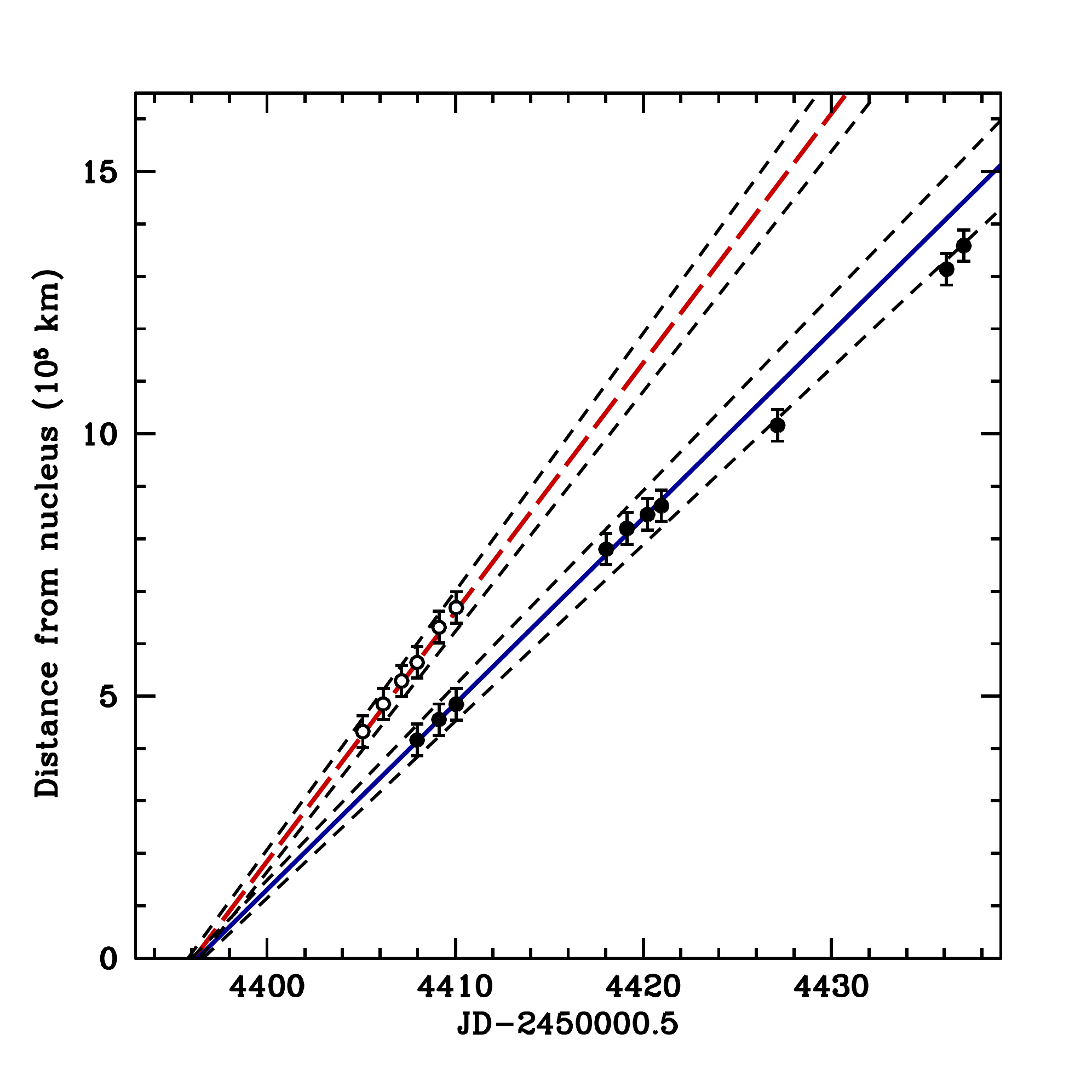}
\caption{\small Distances from the nucleus to the intersection points of
  Regions 1 and 2 (solid circles) and Regions 2 and 3 (open circles) of
  the linear surface brightness profiles of 17P's coma, as
  described in the text, between 2007 October 31-November 1
  and 2007 December 2-3 (where JD-2450000.5=4400.0 corresponds to 2007 October 27.0).
  The best-fit linear function following the movement of the first
  intersection point is shown as a solid line, while the best-fit
  linear function following the movement
  of the second intersection point is shown as a large-dashed line.
  The last three intersection points of Regions 1 and 2 are plotted
  for reference but are not included in the fit as explained in \S\ref{comaexp}.
  Small-dashed lines indicate the range of linear fits permitted by the
  parameter uncertainties we report for the best-fit functions.
  }
\label{comaexpansion}
\end{figure}

We define the transition point between two regions of the coma's surface
brightness profile as the intersection of the best-fit linear functions
(in log-log space) of those regions. As we seek to avoid being affected by
fluctuations in sky brightnesses and the steady fading of the outer coma, we
first only concern ourselves with the positions of the intersections of Regions
1 and 2 and Regions 2 and 3 (Table~\ref{expansion}), ignoring intersections
of those regions with Region 4 (the sky). Plotting the distances
of these intersection points from the nucleus as a function of time
(Fig.~\ref{comaexpansion}), we find best-fit linear solutions corresponding
to an expansion velocity of $v_{exp}=0.41\pm0.02$~km~s$^{-1}$ and a
start date of $t_0(1-2)=2007~{\rm October}~23.3\pm0.3$ (very close to
the time of the final SuperWASP observation of the comet on the night prior
to the outburst's discovery)
for the intersection point between Regions 1 and 2. An expansion
velocity of $v_{exp}=0.55\pm0.02$~km~s$^{-1}$ and a start date of
$t_0(2-3)=2007~{\rm October}~23.1\pm0.3$) is found for the
intersection point between Regions 2 and 3.  
The latter expansion rate is consistent with previously reported
values \citep[e.g.,][]{sno07,gai07,lin09}
We believe $t_0(1-2)$
to be the more reliable extrapolated value as it is derived using a reference
point within the dust coma itself.  By contrast, $t_0(2-3)$ is derived
using the intersection of the dust-dominated and gas-dominated
components of the coma.

For the solution for the outburst time using the intersection between Regions 1
and 2, we use data up to November 17 only.  Beyond this date, the location of
this reference point within the expanding frame of the coma appears to have
shifted as expected (discussed above), and as such, we omit these data in
this portion of our analysis.

For completeness, we also consider the intersections of Region 2 and
Region 3 with the sky (Region 4).  Plotting the distances of these points
(Table~\ref{expansion}) as a function of time 
and omitting data points from 2007 Oct 31-Nov 1 (due to the moderately
high sky background on that night) and from 2007 Nov 5-6 onwards (due to
the notable deviation of points from those nights from the trend exhibited
by points prior to 2007 Nov 5-6), we find best-fit linear solutions
corresponding to an expansion velocity of $v_{exp}=1.7\pm0.1$~km~s$^{-1}$
and a start date of $t_0(2-4)=2007~{\rm October}~22.8\pm0.5$ for
the intersection point between Region 2 and the sky. For the intersection
point between Region 3 and the sky we measure an expansion
velocity of $v_{exp}=0.4\pm0.1$~km~s$^{-1}$ and a start date of
$t_0(3-4)=2007~{\rm October}~17.6\pm0.5$.  These derived expansion
start dates for these reference points are significantly earlier than
plausible from reported observations, which is the expected result of using
a method that  underestimates the coma's expansion velocity.

\subsection{Kinematic Structure of the Coma}

In \S\ref{comaexp}, we find that outburst times, $t_0(1-2)$ and $t_0(2-3)$,
as derived from tracking material at two different locations
in 17P's outer coma were coincident, strongly suggesting that
this part of the coma is dominated by material ejected in a near-instantaneous
explosive manner, rather than over an extended period of time.
Specifically, by fitting the projected velocity $v_{exp}$ of material
in the outer coma at a given projected distance $r$ from  the nucleus
on a given date $t$, we find the following relationship:
\begin{equation}
v_{exp} = 0.95 {r\over t-t_0} \pm 0.02~{\rm km~s}^{-1}
\label{comavelocity}
\end{equation}

This velocity gradient is most likely a particle size effect,
where smaller particles have larger
velocities imparted to them by gas drag during their initial ejection
due to their smaller ratio of mass to surface area.
The observed projected velocity of an ejected dust particle also depends
on the angle of its ejection from the nucleus with respect to the plane
of the sky, where dust particles with velocity vectors perpendicular to
the line of sight ({\it i.e.}, parallel to the plane of the sky) will have
the greatest apparent velocities.  Thus, while all material observed at a
given point in the coma is moving at the same projected velocity from the
nucleus, that material actually consists of a range of highly-projected,
small, fast-moving particles, and less-projected, larger, more
slowly-moving particles.  These particles are additionally affected by
solar radiation pressure which eventually causes particles having identical
projected ejection velocities but different sizes to diverge, as can be seen
by the evolution of the shape
of 17P's coma profile ({\it cf}. Table~\ref{lsbparams}; Fig.~\ref{lsbprofiles}).
Given the large ejection speeds of particles in the outer coma
relative to the acceleration due to solar radiation, the overall effect
remains relatively small over the short time interval that we consider here.

Given the low
spatial resolution and saturation
in much of our SuperWASP data, we are unable to probe the comet's
inner coma in sufficient detail to rule out the possibility of later
emission from the nucleus.  \citet{mon08} did however perform an analysis
of the inner coma, fitting a bidimensional Gaussian to the innermost portion
of the dust cloud ($\sim30$~arcsec), centred on the dust cloud's brightness
peak (instead of on the nucleus as in our analysis), and plotting the mean
of the $\sigma_x$ and $\sigma_y$ of their best-fit solutions as a
function of JD.   They derived an average expansion velocity of
$v_{exp}=0.200\pm0.004$~km~s$^{-1}$ for material located approximately
$1.05\times10^5$~km from the cloud centre on 2007 November 02,
and a radial velocity gradient of
$v_{exp} = r \cdot (0.3\pm0.2)\times10^{-5}$~s$^{-1}$ on the same date.
Given the stated uncertainties, their derived velocity gradient is consistent
with the velocity gradient we expect from Equation~\ref{comavelocity}
($v_{exp} = r \cdot (0.13\pm0.02)\times10^{-5}$~s$^{-1}$), although
their derived velocity ($v_{exp}=0.200\pm0.004$~km~s$^{-1}$)
for material at $1.05\times10^5$~km from the
cloud centre is not consistent
($v_{exp}=0.14\pm0.02$~km~s$^{-1}$,
according to Equation~\ref{comavelocity}).

The large plate scale of SuperWASP's CCDs means that we
cannot probe the central 30~arcsec of 17P's coma.  Beyond this distance, the coma
deviates significantly from a Gaussian profile
\citep[as also noted by][]{mon08}.  As such, we are unable to repeat
and verify the analysis described above using our data.  We do note, however, that the
large amount of scatter in their data and the lack of actual convergence at a
single start time of their plots showing the expansion of the coma in four
different filters suggests that their value for the velocity gradient in the
inner coma may be unreliable.
We also note that their computed expansion velocity and extrapolation of their
coma expansion plots imply ejection of the material on 2007 October $27.0\pm0.7$,
significantly later than the actual
observed start of the outburst on the night of 2007 October 23-24.
This large discrepancy between implied ejection
times, even accounting for uncertainties, strongly suggests that the
material in the inner coma at the time of their observations
was not in fact ejected during the initial
outburst.  In turn, this suggests the existence of some continued
emission from the nucleus as late as 2.5 days following the initial outburst.

\subsection{The ``False Nucleus''\label{nuclfrag}}

As noted by many observers of 17P, the comet's coma actually exhibited
two brightness peaks, one of which was the comet's nucleus and another which
we hereafter refer to as the ``false nucleus.''  This second
brightness peak was also referred to in observation reports as a ``chunk''
or ``blob'', and was 
hypothesised by some to be due to a large nuclear fragment that had
detached from the primary comet nucleus and was slowly
drifting away \citep[{\it e.g.},][]{sno07,tri07,dra07}.
Arai et al. (using data from 2007 October 27 and October 30) and
Montalto et al. (using data from 2007 October 27 and October 31),
measured projected drift velocities of 150~m~s$^{-1}$ and
127~m~s$^{-1}$, respectively \citep{tri07}.  Montalto et al. additionally
derived a separation time of 2007 Oct $25.6\pm0.2$, and later
reported a revised drift velocity of $135\pm1$~m~s$^{-1}$ \citep{mon08}.
Subsequent analysis of 17P images by investigators such as Lecacheux et al.
\citep[in][]{tri07}, \citet{ste09}, and \citet{rea10} revealed the
false nucleus to be concentrations of material in a series of dust
trails receding from the nucleus.

\begin{figure}
\includegraphics[width=5.5in]{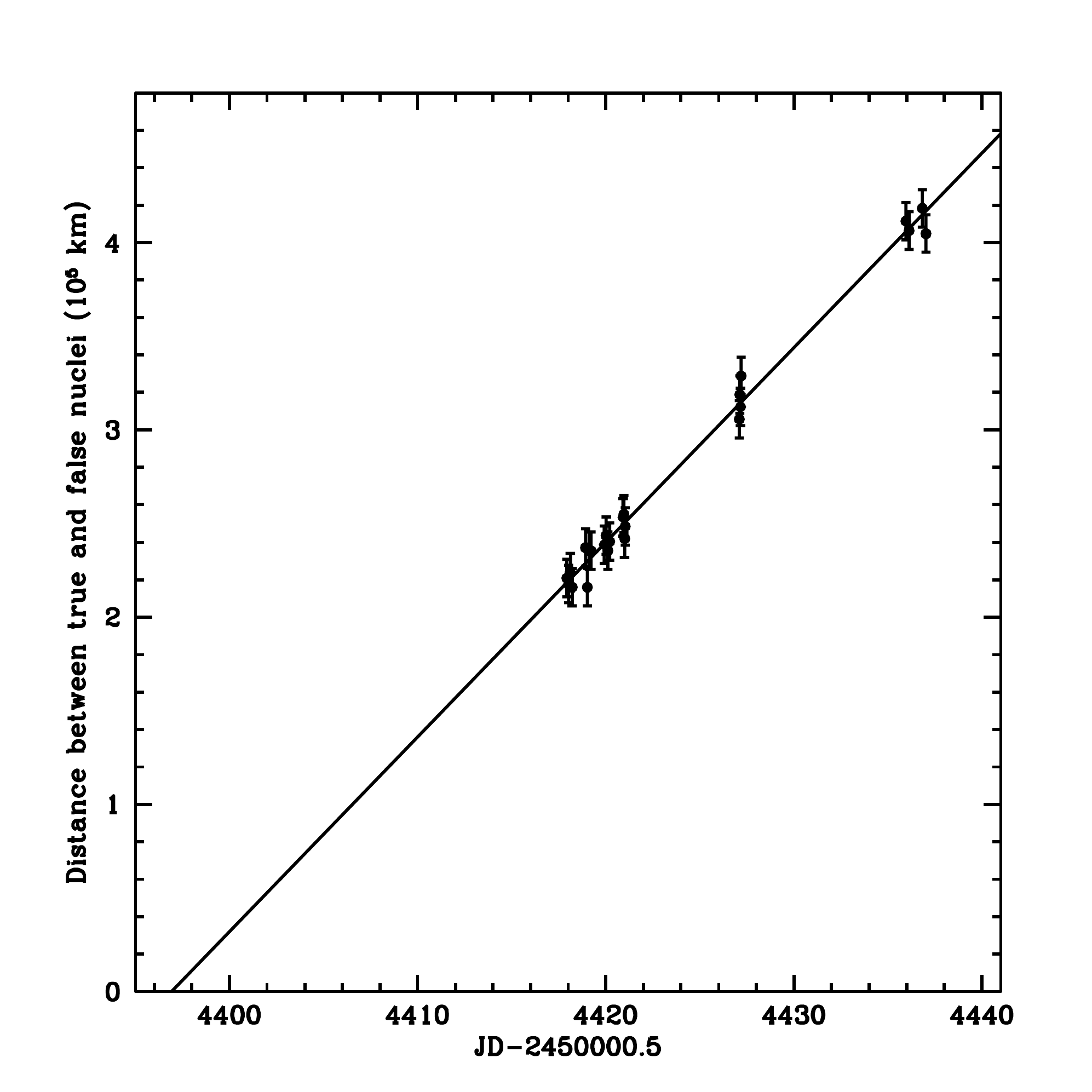}
\caption{\small Distance (solid circles) between the nucleus and separated
  fragment of 17P between 2007 November 5-6
  and 2007 December 2-3 (where JD-2450000.5=4400.0 corresponds to 2007 October 27.0), plotted as a function of time.
  The best linear fit is shown as a solid line.
  }
\label{chunkdist}
\end{figure}

To test reported velocity measurements of the densest concentration of
material in these dust trails, we measure projected distances between
the true nucleus and false nucleus in our own data taken between 2007 Nov 5 and
2007 Dec 3 (Table~\ref{fragmotion}) and plot these as a function of time
(Fig.~\ref{chunkdist}).  Assuming a constant drift velocity, we find a best-fit
projected linear velocity of $v_{drift}=120\pm5$~m~s$^{-1}$ and an initial
separation time of $t_{sep}=2007~{\rm October}~23.9\pm0.5$, roughly
consistent with other derived outburst times.  The
estimated uncertainties primarily reflect the difficulty of precisely
determining the position of the rather ill-defined false nucleus on each night.

\subsection{Dust Modelling\label{dustmodel}}

One key to understanding the nature of 17P's outburst is determining
whether the comet's nucleus is a plausible primary emission source,
as we have assumed thus far, or if the coma could be dominated by
dust ejected from another source, perhaps a separated fragment, as
others \citep[{\it e.g.},][]{sek08a},
have suggested as a possibility.  With this question in mind, we
construct a simple dust model aimed at constraining the possible source
(or sources) of emission.
We make several simplifying assumptions, as in any model, and do not
seek a complete analytical description of the outburst event.
Instead, we simply aim to determine whether 17P's nucleus is a
plausible origin of the dust in the outer coma, given its significant
offset from the coma centre (and the approximate location of the false
nucleus) in observations two weeks after the initial
outburst and later.

We define the plane of the sky as the $x$-$y$ plane, where
the comet nucleus is located at the origin and the $x$-axis is
perpendicular to the projection of the sunward vector
(the positive $y$-axis) in the plane of the sky.
The $z$-axis is taken to be aligned with the line of sight,
with the positive axis extending towards the observer.

Assuming a simplistic model of isotropic emission in which particles are
ejected at terminal ejection velocities of $v_{ej}$ in all directions,
we expect the coma to be initially spherical and then become
paraboloidal as solar radiation pressure applies a uniformly-directed
acceleration, $a_{rp}$, to each coma particle \citep[{\it cf}.][]{edd10},
Following convention, we parametrize particle size by the dimensionless
parameter, $\beta$, which is defined as the ratio of a particle's acceleration
due to radiation pressure to its acceleration due to solar gravity.
Formally, $\beta$ is a complicated function of the size, composition,
and shape of a given dust grain \citep{bur79}, though in practice,
it can be approximated by $\beta\approx1/a_d$, where $a_d$ is the
radius of a given dust particle in $\mu$m.

Using the $\beta$ parametrization, the acceleration of a particle
due to solar radiation pressure is then approximately given by
\begin{equation}
a_{rp} = \beta {g_{\odot}\over R_{\rm AU}^2}
\label{radpress}
\end{equation}
where $g_{\odot}=0.006$~m~s$^{-2}$ is the gravitational acceleration to
the Sun at 1 AU and $R_{\rm AU}$ is the heliocentric distance in AU.

We adopt a typical particle size-velocity relationship of
\begin{equation}
v_{ej} \sim v_0\beta^{1/2}
\end{equation}
where $v_0$ is the reference ejection velocity in m~s$^{-1}$ of a
particle with $\beta=1$ \citep[{\it cf}.][]{ish07,hsi09}.

Noting the short distance that the comet traveled around its orbit
(from $\nu=61.3^{\circ}$ to $\nu=65.0^{\circ}$; Table~\ref{obslog})
during the first 15 days of the outburst (from 2007 October 23.6 to
2007 November 5-6; the period to which we confine our modelling
analysis), we make the simplifying approximation of
treating the orientation of the solar radiation pressure acceleration
vector as constant during this period.  The $(x,y,z)$ coordinates of
a dust particle, as viewed at opposition, ejected from the nucleus
at an elevation angle of $\theta$ and azimuth angle of $\phi$, are
then given by
\begin{equation}
x = v_{ej} t \cos(\theta)\cos(\phi)
\end{equation}
\begin{equation}
y = v_{ej} t \cos(\theta)\sin(\phi)
\end{equation}
\begin{equation}
z = -{1\over 2}a_{rp} t^2+v_{ej} t \sin(\theta)
\end{equation}
which we calculate for $-90^{\circ}\leq\theta\leq90^{\circ}$ and
$0^{\circ}\leq\phi<360^{\circ}$ in $1^{\circ}$ intervals for both
$\theta$ and $\phi$.
Then, rotating our coordinate
system to account for a non-zero phase angle, $\alpha$, at the
time of observation, we obtain
\begin{equation}
x' = x
\end{equation}
\begin{equation}
y' = y \cos(\alpha) + z \sin(\alpha)
\end{equation}
for the $(x',y')$ coordinates (in physical distance) of our test
dust particles in the plane of the sky.  These coordinates are then
divided by the apparent pixel scale to obtain our final $(x',y')$
coordinates in pixels.

In this modelling study, our focus is on
the fastest-moving particles in the outer envelope of the dust coma,
and we are therefore only concerned with the minimum
observed particle size in the coma, or $\beta_{max}$.
Thus, using the geometric circumstances of 17P on
2007 November 5-6, we generate a series of model dust clouds using
various values for $\beta$ and $v_0$ that satisfy
$v_{ej}=550$~m~s$^{-1}$ ({\it i.e.}, the velocity we derive for particles
on the outermost edge of the dust coma).

We find that each of our parameter sets selected in this way produces a nearly
circular dust cloud (as projected in the plane of the sky)
with a diameter of about $18'$, consistent with
the actual size of the dust envelope on 2007 November 5-6 (as expected,
given the constraints we applied to our initial parameters).
As larger $\beta$ and smaller $v_0$ values
are used, the dust cloud remains fixed in size but
becomes increasingly offset from the point of emission.
Physically, the fixed size of the cloud results from our
constraint on $v_{ej}$ described above,
while the shifting position of the cloud as $\beta$ increases
corresponds to smaller particles
experiencing larger accelerations due to solar radiation pressure.
The cloud therefore ``drifts'' more quickly relative to the nucleus.
In the limiting case of particles with very small
$\beta$ values ({\it i.e.}, very large particles), solar radiation pressure
would not cause the dust to drift at all, and the nucleus
would therefore appear to be at the exact centre of the dust cloud.

\begin{figure}
\includegraphics[width=5.5in]{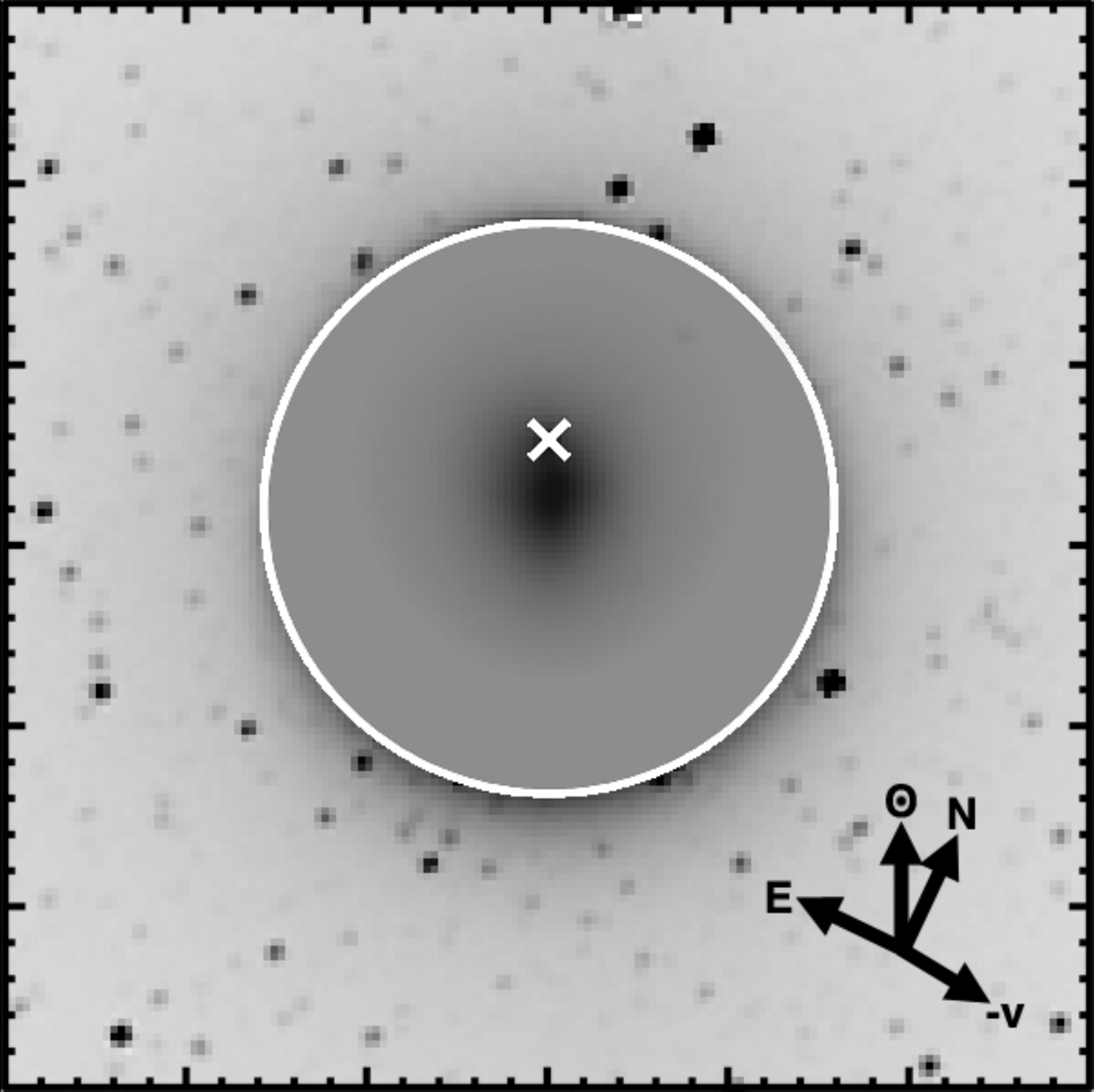}
\caption{\small A SuperWASP image of 17P from 2007 November 5-6
with the outer envelope of a modelled dust cloud, marked with a solid white line, overlaid.
Different contrast scaling is used for the comet and the sky to emphasise detail in the inner coma. 
The model dust cloud consists of
particles sharing a single $\beta$ value ejected in a single impulsive event.
The origin point of modelled dust emission is
marked with an ``X'' and coincides with the image of 17P's nucleus in the
observed data. The modelled dust envelope was generated using $\beta=1.0$,
$v_0=550$~m~s$^{-1}$, $R=2.49$~AU, $\Delta=1.62$~AU, $\alpha=13.8^{\circ}$,
and a pixel scale conversion of $1.61\times10^4$~km~pix$^{-1}$ (corresponding
to the SuperWASP pixel scale of $13\farcs7$~pix$^{-1}$), and is coincident with
the outer envelope of the observed dust coma.
The image shown is approximately $35'$ by $35'$ and is oriented such that the
sunward vector (as projected in the sky) corresponds to the positive vertical axis.
}
\label{modeloverlay}
\end{figure}

Assuming 17P's nucleus to be the primary origin point of the
particles in the outer dust coma, we find that the model dust cloud that
best approximates the relative positions of the nucleus and the
observed outer edge of the coma is produced by
particles with $\beta=1.0$ (or $a_d\sim1~\mu$m) and $v_0=550$~m~s$^{-1}$
(Fig.~\ref{modeloverlay}).  We note that these model parameters are selected
to match our observed expansion rate, and as such, model
coma sizes and measured coma sizes agree by definition.
The model dust envelopes plotted in Figure~\ref{modeloverlay} are produced by
isotropic emission as described above, but we also note that identical
dust envelopes can also be produced by semi-isotropic ({\it i.e.}, hemispherical)
emission solely from
either the sun-facing hemisphere or the non-sun-facing hemisphere.  This
finding is consistent with \citet{mor08} who found that the outer shell of
the dust coma could be fit by emission from the entire sun-facing
hemisphere of 17P.

As stated above, the simplicity of our model does not allow us to
comprehensively explain all aspects of 17P's dust morphology.
Nonetheless, we show that, for reasonable dust particle sizes
($a_d\gtrsim1~\mu$m), the position of 17P's
dust coma relative to the nucleus in the two weeks after
the start of the outburst can be accounted for by a single
isotropic (or semi-isotropic) emission event where 17P's nucleus is
the source of emission.
In contrast, particles in a dust cloud for which, for example, a large
separated fragment with the same motion as the false nucleus
({\it cf}. \S\ref{nuclfrag}) is the emission origin
({\it i.e.}, where the cloud shows minimal drift relative to its source)
would have to have
$\beta_{max}=0.5$ ($a_d\gtrsim2~\mu$m; where $v_0=780$~m~s$^{-1}$).

In terms of particle size constraints,
both scenarios are essentially equally plausible, though we note that a more
powerful ejection mechanism (to account for the larger value of $v_0$)
is required in the second scenario where emission originates from the false
nucleus.  Admittedly,
even the relatively more modest $v_0=550$~m~s$^{-1}$ value required for the
true nucleus to be the emission source is high,
though it is at least not unprecedented;  the 1836 outburst
of 1P/Halley was characterised by a coma expansion rate of 575~m~s$^{-1}$ \citep{sek08b}.
In terms of outburst mechanics, however, the former scenario is must be considered more
likely as it only requires a single explosive outburst taking place on 17P's nucleus, whereas
the latter scenario requires a two-part event where a large fragment first detaches at
high speed (projected velocity of $\sim$120~m~s$^{-1}$; \S\ref{nuclfrag}) from the primary nucleus and then subsequently explosively disintegrates.

\subsection{Gas Modelling\label{gasmodel}}

\citet{sek08a} and others have cautioned that 17P's dust coma could
be significantly contaminated by the gas coma.  
\citet{kob07} observed
gas emission bands on 2007 October 25.46, but not on 2007 October 24.58,
suggesting that ice grains ejected in the initial outburst did not
sublimate until some point in the interim period between their observations.
Wagner et al. were
able to trace CN (0-0) emission as far as $2\farcm4$ from the photocentre
of the dust cloud in observations taken on 2007 October 25.181 and 25.465
\citep{kob07}, implying an average velocity of 1.6~km~s$^{-1}$ to
1.9~km~s$^{-1}$, consistent with our estimated gas coma expansion
velocity of $\gtrsim$1.7~km~s$^{-1}$ (\S\ref{comaexp}). \citet{sch09} observed steadily
decreasing production rates of several gas species between 2007 November 1 and
2008 March 5.

In terms of our data, it is possible that the portion of 17P's coma surface
brightness profile designated Region 3 in our analysis simply fades below the
background detection limit.  However,
we hypothesise that this region in fact
corresponds to the comet's gas coma, and that by 2007 November 13-14, this
region had disappeared from the coma surface brightness profile because
most of the gas molecules initially released had photodissociated or ionised and the
remaining coma was below our detection limit. 

To test this hypothesis, we construct a simple model of
17P's gas coma based on the Monte-Carlo model of \citet{com80} but with
additional modifications. First, we assume that
the parents have an isotropic Maxwellian velocity distribution
with a mean velocity of  $0.54$ km/sec. This would be expected for a
normal comet undergoing steady-state sublimation \citep{coc86}, though it is
possible that this velocity is an underestimate since an explosive release of
material could result in higher acceleration of gas from the nucleus
than usual. In addition, we have adapted the model to assume either a
single outburst in terms of a delta function for the production of gas,
or an outburst where the production of parent molecules instantaneously
starts at a maximum value and thereafter falls off exponentially.
Neither scheme may be accurate in detail, but they allow us to conduct
a basic analysis of gross gas coma morphology. 

Gas emission in the SuperWASP images was likely dominated by the
commonly observed (0-0) band of C$_2$.  The production of this species
has been investigated by many investigators due to its domination of the
optical emission spectrum of comets. A recent in-depth study of C$_2$
chemistry in comet Hale-Bopp by \citet{hel05} concluded that C$_2$H$_2$
was the primary parent molecule, but that C$_2$H$_6$ was an additional
minor parent and electron impact dissociation is important to consider
as well as photodissociation.  A model including multiple production and
destruction mechanisms that properly accounts for relative reaction rates
is beyond the scope of this
paper.  Instead, we adopt a simple two-step single parent photodissociation
model as used in the past for other comets. Although \citet{hel05} show
such a model is not physically accurate, previous
authors have been able to fit C$_2$ coma profiles with such a scheme by
using suitable parameters.
Here we assume the photodissociation values
found by \citet{led09} of
a parent lifetime of $3.4\times10^4 R^2$ s, a C$_2$ daughter lifetime
of $1.2\times10^5 R^2$ s, and an excess C$_2$ velocity from dissociation
of 1.2~km~s$^{-1}$.

\begin{figure}
\includegraphics[width=5.5in]{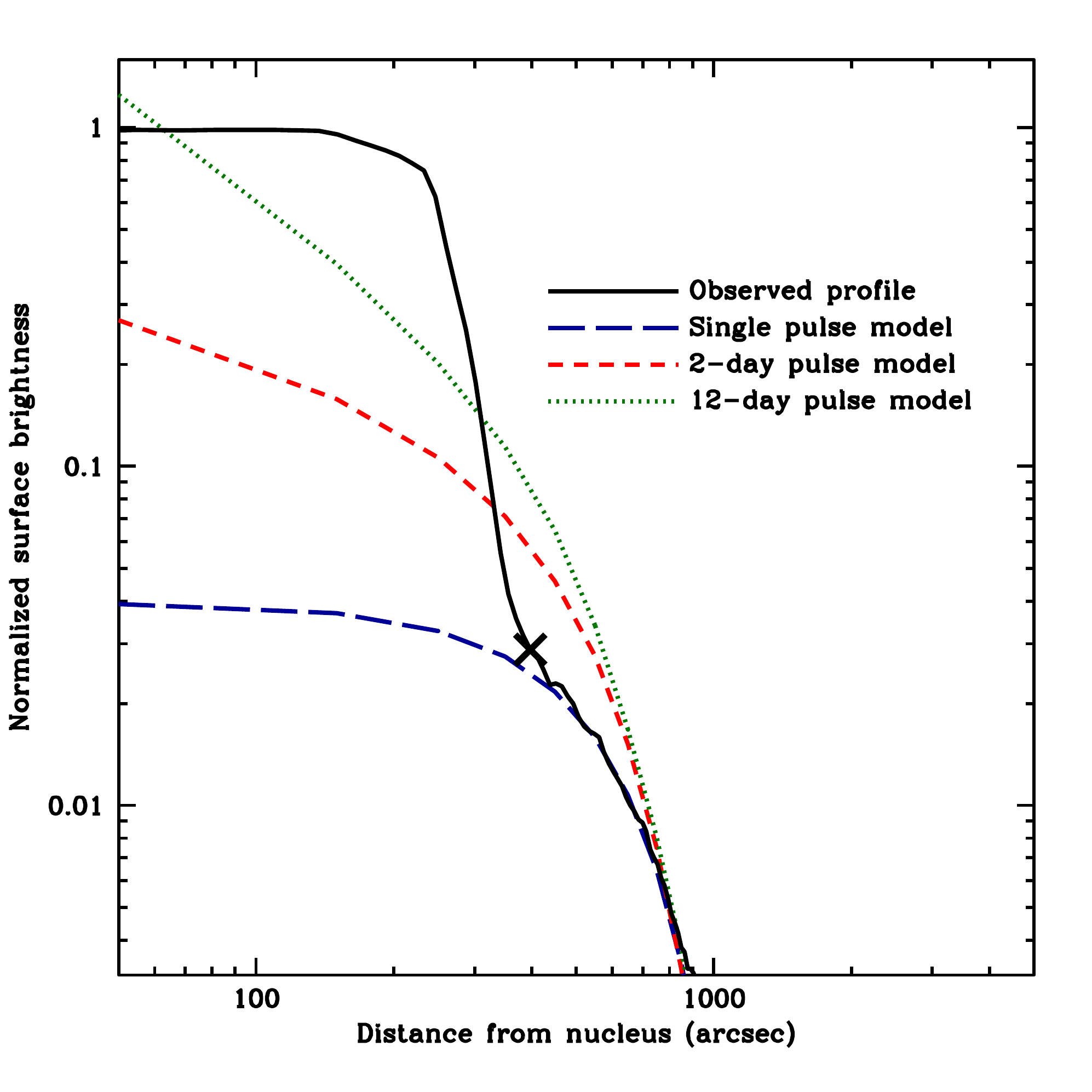}
\caption{\small Comparison of the observed surface brightness profile of
the coma of 17P (solid line) on the night of 2007 October 31 - November 1
({\it cf}. Fig.~\ref{lsbprofiles}a),
and plots of modelled C$_2$ column density. The point where the observed coma
transitions from dust-dominated to gas-dominated is marked by ``X''.
Large-dashed line: Precursor molecules are all released from the nucleus in a single
pulse on 2007 October 23.6.
Small-dashed line: The release of
precursor molecules begins on 2007 October 23.6 but then exponentially
decreases with a 48-hour timescale.
Dotted line: As above with the gas production exponentially
decreasing with a 12-day timescale.
Model profiles have been scaled vertically
to produce the best possible fit to the observed coma profile.
}
\label{gasplot}
\end{figure}

In the model we generated $10^9$ molecules and calculated the resulting
surface brightness profiles for a pulse outburst, one with a two-day
exponential decay time, and one with a 12-day lifetime
as suggested from latetime gas measurements by \cite{sch09} (we note that
Schleicher found a faster rate of decay in activity than an exponential).
In Figure~\ref{gasplot}, we overplot these model profiles on 17P's observed surface brightness
profile for the night of 2007 October 31 - November 1, scaling the model plots
vertically (i.e., varying the column density of the modelled gas
emission) in order to produce the best possible fit to the observed
surface brightness profile in the gas-dominated region.
As can be seen, we find that the single pulse
model offers an excellent fit to the Region 3 portion of the coma surface
brightness profile.  By contrast, the suitability of the two-day pulse model is 
considerably worse as it produces a profile that is too
steep, regardless of scaling, to match the observed profile. The 12-day
pulse model produces an even steeper profile and, therefore, an even worse fit.

Taking the single-pulse model as best fitting the outer coma as observed
in our images, we note that the expected lifetime of the C$_2$ coma
generated in the initial outburst event is $\sim 10^6$~s or $\sim13$ days.
This expected lifetime is consistent with an
outer gas coma that is visible as Region 3
in our images up to 2007 Nov 5-6 (14 days after outburst), but is absent
in images obtained from 2007 Nov 13-14 onwards (22 days after outburst).

\section{DISCUSSION AND SUMMARY}

\subsection{Primary Findings}

Wide-field imaging of the 2007 outburst of 17P/Holmes was
serendipitously obtained by the SuperWASP-North
facility on 17 nights over a 42-night period
beginning on the night (2007 October 22-23) immediately prior to the
outburst.  We report the following key findings:
\begin{itemize}
\item{The comet was not detected in data from the night
  before the outburst's discovery on the night of 2007 October 22-23,
  indicating that it remained below the SuperWASP detection threshold of
  $m_V\sim15$~mag (consistent with pre-outburst reports placing the comet's
  brightness at $m_V\sim17$~mag) until at least 2007 October 23.3.
}
\item{The unresolved coma (as seen by SuperWASP)
  was likely optically thin during our observations on the first
  night of the outburst.  The comet's lightcurve during
  these observations is consistent with an exponential function,
  suggesting that its rapid brightening could have been driven by the
  progressive fragmentation of ejected material produced on
  a very short timescale at the time of the initial outburst. 
  Our best-fit functions to the data imply an initial
  fragmentation timescale of $t_{frag}\sim2\times10^3$~s, decreasing to
  $t_{frag}\sim1\times10^3$~s,
  and a near-instantaneous leap in brightness from a magnitude of
  $m_V\sim17$~mag to $m_V\sim11$~mag at
  the moment of the initial outburst.
}
\item{Analysis of coma surface brightness profiles reveals a velocity
  gradient consistent with the outer coma
  being dominated by material ejected in an explosive manner (i.e., at
  a single instant), rather than
  over an extended period of time.  Near the outer edge of the visible coma,
  where the coma likely transitions from being dust-dominated to gas-dominated,
  this velocity gradient corresponds to an expansion velocity of $0.55\pm0.02$~km~s$^{-1}$,
  consistent with previous reported measurements.  From this analysis, we
  find a most likely outburst date of $t_0=2007~{\rm October}~23.3\pm0.3$.
  This finding of explosively ejected material dominating the outer coma does
  not rule out the possibility of subsequent sustained mass ejection supplying 
  the inner coma at later times.
}
\item{We measure the rate of motion (relative to the nucleus)
  of the secondary brightness peak in 17P's coma, which we
  refer to as the ``false nucleus'', and find a projected drift velocity of
  $v_{drift}=120\pm5$~m~s$^{-1}$, consistent with previously reported
  measurements.
}
\item{Dust modelling shows that 17P's nucleus is
  a plausible primary emission source of outer coma material, and that a secondary
  source such as a separated nucleus fragment is not required to explain the
  motion of the coma relative to the nucleus.
  We show instead that the drifting of the coma relative to the nucleus can be
  explained as the consequence of radiation pressure alone.
}
\item{Modelling of 17P's gas coma indicates that the morphology of the portion of
  the observed coma profile hypothesised to be gas extending past the outer dust coma
  is best explained by a single, instantaneous outburst of gas production,
  rather than extended gas production that persists over several days.
  This result is consistent with our conclusion that the outer dust coma is likely
  dominated by material ejected instantaneously, and not over an extended period of time.
  We also note that C$_2$ is likely to be the dominant observed component of
  the gas coma, and find that its decay time is consistent with SuperWASP observations
  of the disappearance of the observed coma profile region attributed to gas.
}
\end{itemize}

\subsection{Anatomy of an Outburst}

To date, no consensus has emerged to explain the physical origin of
17P's spectacular outbursts in 1892 and 2007.  A primary difficulty lies in
reconciling the occurrence of such an apparently catastrophic event at least
twice in the recent past with a lack of comparable outbursts
in almost all other comets, although \citet{sek08b} has noted that 1P/Halley
exhibited a similarly powerful outburst in 1836.
\citet{whi84} suggested that a grazing
encounter and eventual impact by a small satellite with the nucleus could
have been responsible for the 1892 outburst, but such an explanation was
rendered highly implausible with a second episode of outburst activity
in 2007. Furthermore, any model formulated to explain 17P's behaviour must
also account for the lack of outburst activity in the 115 years between
1892 and 2007 \citep{sek09b}.

\citet{sek08a,sek08b} has proposed a scenario in which
an inwardly-diffusing thermal wave gradually penetrated a large,
weakly-cemented pancake-shaped layer of the nucleus \citep[analogous to
those observed for 9P/Tempel 1; {\it cf}.][]{tho07} over numerous orbits
until it finally reached a large reservoir of amorphous water ice at the
layer's base. Upon being heated to
the necessary transition temperature, this
amorphous ice layer underwent an exothermic transformation to crystalline
ice, leading to rapid sublimation and causing the pancake-shaped
layer to separate from the nucleus and almost immediately
explosively disintegrate.  This two-part outburst scenario is invoked
by Sekanina to explain the appearance of the false nucleus as
the central source of 17P's dust coma, as well as its apparent motion
relative to the true nucleus.

We note, however,
that our dust modelling (\S\ref{dustmodel}) shows that it is not necessary
for the false nucleus to be the source of emission to explain the position
of the outer dust coma relative to the true nucleus.  Thus a model where a
large surface fragment first violently separates from the nucleus
(with a projected velocity exceeding 100~m~s$^{-1}$) and then explosively
disintegrates, requiring two non-simultaneous catastrophic events,
may be needlessly complicated.  The false nucleus could still
be a concentration of material related to a large fragment ejected in
the initial outburst that then disintegrated, but may be better
characterised as a byproduct, rather than a key component, of the outburst.

\citet{alt09} propose an alternate hypothesis in which a particularly
close perihelion passage by 17P on 2007 May 4 substantially raised the
sublimation rate of subsurface water ice relative to previous perihelion
passages.  This increased production of subsurface water vapour, coupled with an
``airtight'' layer of surface regolith, ultimately led to an explosive
disintegration of 17P's dust mantle 172 days later, causing the observed outburst.
While it is true that 17P's perihelion passage just 2.05~AU from the Sun was
its closest approach to the Sun in over a century (according to JPL's
Horizons ephemeris generator at http://ssd.jpl.nasa.gov/horizons.cgi),
we note that 17P's perihelion
distance immediately prior to its 1892 outburst was a more modest 2.14~AU
from the Sun, a distance comparable to perihelion passage distances in 1864,
1871, 1878, 1885, 1899, 1906, 1972, 1979, 1986, 1993, and 2000, all of which
were reached without any recorded reports of behaviour comparable to the
comet's 1892 or 2007 outbursts.

Most recently, \citet{rea10} propose a scenario in which trapped gases
released by the crystallisation of amorphous ice, as well as sublimation of
other ices driven by the exothermic crystallisation process, cause a
buildup of gas pressure in a subsurface cavity on the comet.  The outburst
of 17P then occurred when the pressure built up within this cavity exceeded
the strength of the surrounding material (found to be relatively high --- $10-100$~kPa
--- based on the nucleus's survival of the outburst), causing the cavity to rupture
violently and energetically, creating the observed explosive ejection of
nuclear material. 
We note that while \citet{kos10} report that the crystallisation of amorphous water
ice is unlikely to have caused 17P's outburst, they reach this conclusion based
on a simple model in which the ice being crystallised is present in a sub-surface
layer just below a dust mantle.  The hypothesis presented by Reach {\it et al.} is
based on a different physical scenario where ice crystallisation
occurs in a subsurface void in which gas pressure is able to build up, and as such,
we believe that it remains plausible and, 
of the hypotheses proposed thus far, is
the most consistent with our own findings.
We note, however, that it then raises
questions as to why 17P is the only comet known thus far
with the apparently unique combination of high-tensile-strength nuclear material,
subsurface cavities, and amorphous ice capable of driving two of the largest
cometary outbursts observed in modern astronomy.  

\citet{gai07} reported that observations they made between 2007 October 24 to
2007 November 4 with the Pic du Midi 1~m telescope show multiple
dust streams with well-defined origin points, four of which were measured
to recede from 17P's nucleus at roughly constant velocities (as
projected on the sky) ranging from 50 to 100~m~s$^{-1}$, implying the
presence of unresolved, steadily disintegrating nucleus fragments in the coma.
These fragments were also calculated to have separated from the nucleus
between 2007 October 23.7 and 2007 October 24.8.
Additionally, \citet{ste09} observed at least sixteen
10-m- to 100-m-scale fragments receding from the nucleus that showed
evidence of ongoing sublimation and disintegration. Coupled with our finding of dust fragmentation in the
first hours after outburst, these observations point to a scenario of continued fragmentation of cometary material
following the outburst.

We certainly still have far from a complete picture of 17P's 2007 outburst.
We suggest that detailed dust modelling and analysis of images of the comet's
inner coma using data with higher spatial resolution than SuperWASP would
be useful for clarifying the temporal and kinematic nature of the outburst,
in particular whether significant dust production continued from the nucleus
after the initial outburst event and, if so, whether ejection velocities
were comparable to those in the initial outburst event.
More detailed discussions of spectroscopy would also help constrain the spatial and temporal nature of
the contribution of gaseous species to the 17P's appearance, such as whether
sublimation of nucleus-bound ices or ice particles in the coma was
significant, and whether there was any appreciable delay in the sublimation
of icy material in the coma that could be linked to hierarchical fragmentation
of macroscopic nucleus particles.

\section*{Acknowledgements}
We appreciate support of
this work through STFC fellowship grant ST/F011016/1 to HHH.
The WASP consortium comprises scientists
primarily from the University of Cambridge (Wide Field Astronomy Unit), the Instituto
de Astrof\'isica de Canarias, the Isaac Newton Group of Telescopes, the University of Keele,
the University of Leicester, the Open University, Queen's University of Belfast, and the
University of St. Andrews. The SuperWASP cameras were constructed and are operated with
funds made available from the consortium universities and the UK's Science and Technology
Facilities Council.  We also thank an anonymous referee for helpful comments that improved
this manuscript.
This research has made use of the VizieR catalogue access tool, CDS, Strasbourg, France.

\begin{table}
\begin{minipage}[t]{\columnwidth}
\caption{Observation Log}
\label{obslog}
\centering
\renewcommand{\footnoterule}{}
\begin{tabular}{lcrccccccc}
\hline\hline
  UT date range
   & Moon\footnote{Lunar phase expressed in offset from new Moon (``N'') in days}
   & N\footnote{Number of images obtained}
   & $R$\footnote{Heliocentric distance in AU}
   & $\Delta$\footnote{Geocentric distance in AU}
   & $\alpha$\footnote{Solar phase angle (Sun-17P-Earth) in degrees}
   & $\nu$\footnote{True anomaly in degrees}
   & $pa_{-\odot}$\footnote{Position angle in degrees East of North of anti-solar
                   vector as projected on the sky}
   & $pa_{-v}$\footnote{Position angle in degrees East of North of negative
                    velocity vector as projected on the sky}
   & $pa_{tail}$\footnote{Position angle in degrees East of North of central axis
                   of the comet's dust tail, whose direction only becomes
                   clearly identifiable in data from 2007 Nov 3-4 onwards} \\
\hline
2007 Oct 23.0039 -- Oct 23.2659 & N+12 & 72 & 2.43 & 1.64 & 17.3 & 61.1 & 224.8 & 266.0 & --- \\
2007 Oct 23.9851 -- Oct 24.2681 & N+13 & 84 & 2.44 & 1.64 & 17.0 & 61.3 & 223.6 & 265.8 & --- \\
2007 Oct 24.9785 -- Oct 25.2647 & N+14 & 59 & 2.44 & 1.63 & 16.8 & 61.6 & 222.3 & 265.7 & --- \\
2007 Oct 25.9769 -- Oct 26.1961 & N+15 & 31 & 2.44 & 1.63 & 16.5 & 61.9 & 221.0 & 265.5 & --- \\
2007 Oct 31.9649 -- Nov 01.2547 & N--9 & 78 & 2.47 & 1.62 & 15.0 & 63.6 & 212.2 & 264.3 & --- \\
2007 Nov 01.9575 -- Nov 02.2510 & N--8 & 76 & 2.47 & 1.62 & 14.7 & 63.9 & 210.5 & 264.0 & --- \\ 
2007 Nov 02.9518 -- Nov 03.2298 & N--7 & 76 & 2.48 & 1.62 & 14.5 & 64.1 & 208.9 & 263.8 & --- \\ 
2007 Nov 03.9632 -- Nov 03.9825 & N--6 &  6 & 2.48 & 1.62 & 14.3 & 64.4 & 207.2 & 263.5 & 208 \\ 
2007 Nov 04.9503 -- Nov 05.2319 & N--5 & 78 & 2.48 & 1.62 & 14.0 & 64.7 & 205.4 & 263.3 & 208 \\ 
2007 Nov 05.9481 -- Nov 06.2293 & N--4 & 78 & 2.49 & 1.62 & 13.8 & 65.0 & 203.6 & 263.0 & 208 \\ 
2007 Nov 13.9258 -- Nov 14.2159 & N+4  & 76 & 2.52 & 1.63 & 12.2 & 67.1 & 186.9 & 260.8 & 200 \\ 
2007 Nov 14.9262 -- Nov 15.2142 & N+5  & 76 & 2.52 & 1.63 & 12.1 & 67.4 & 184.6 & 260.5 & 200 \\ 
2007 Nov 15.9275 -- Nov 16.2118 & N+6  & 76 & 2.53 & 1.63 & 11.9 & 67.7 & 182.3 & 260.2 & 200 \\ 
2007 Nov 16.9254 -- Nov 17.0383 & N+7  & 30 & 2.53 & 1.63 & 11.8 & 67.9 & 179.9 & 259.9 & 200 \\ 
2007 Nov 22.9603 -- Nov 23.6916 & N+13 & 78 & 2.55 & 1.65 & 11.3 & 69.5 & 164.8 & 258.2 & 195 \\ 
2007 Dec 01.9561 -- Dec 02.1222 & N--8 & 42 & 2.59 & 1.70 & 11.4 & 71.8 & 142.2 & 255.8 & 177 \\ 
2007 Dec 02.8300 -- Dec 03.0240 & N--7 & 60 & 2.59 & 1.70 & 11.5 & 72.0 & 140.2 & 255.6 & 177 \\ 
\hline
\end{tabular}
\end{minipage}
\end{table}

\begin{table}
\begin{minipage}[t]{\columnwidth}
\caption{Field Stars Used for Photometry Calibration}
\label{fieldstars}
\centering
\renewcommand{\footnoterule}{}
\begin{tabular}{lcccr}
\hline\hline
  Star
   & R.A.\footnote{Right Ascension, in hours, minutes, and seconds}
   & Dec.\footnote{Declination, in degrees, arcminutes, and arcseconds}
   & Type\footnote{Spectral type}
   & $m_V$\footnote{$V$-band magnitude} \\
\hline
HIP 16880 & 03:37:13.97 & +49:33:27.12 & F6IV-V & 10.06 \\
BD+48 965 & 03:38:51.74 & +49:24:18.25 & K0     &  9.05 \\
HIP 16965 & 03:38:15.41 & +51:35:22.47 & F4IV   & 10.32 \\
\hline
\end{tabular}
\end{minipage}
\end{table}

\begin{table}
\begin{minipage}[t]{\columnwidth}
\caption{Nucleus Photometry}
\label{photometry}
\centering
\renewcommand{\footnoterule}{}
\begin{tabular}{cccc}
\hline\hline
 UT Date
   & Airmass
   & Flux\footnote{Calibrated net flux of comet nucleus in $10^5$~ADU}
   & $m_V$\footnote{Approximate equivalent V-band magnitude of comet nucleus}
 \\
\hline
2007 Oct 23.9851 & 1.308 & 1.20$\pm$0.07 & 9.67$\pm$0.20 \\
2007 Oct 23.9855 & 1.306 & 1.20$\pm$0.07 & 9.66$\pm$0.20 \\
2007 Oct 23.9938 & 1.271 & 1.29$\pm$0.07 & 9.59$\pm$0.20 \\
2007 Oct 23.9942 & 1.269 & 1.29$\pm$0.07 & 9.60$\pm$0.20 \\
2007 Oct 24.0025 & 1.237 & 1.37$\pm$0.07 & 9.53$\pm$0.20 \\
2007 Oct 24.0029 & 1.236 & 1.38$\pm$0.07 & 9.52$\pm$0.20 \\
2007 Oct 24.0113 & 1.207 & 1.44$\pm$0.07 & 9.47$\pm$0.20 \\
2007 Oct 24.0117 & 1.206 & 1.43$\pm$0.07 & 9.48$\pm$0.20 \\
2007 Oct 24.0201 & 1.181 & 1.53$\pm$0.07 & 9.42$\pm$0.20 \\
2007 Oct 24.0205 & 1.180 & 1.55$\pm$0.07 & 9.40$\pm$0.20 \\
2007 Oct 24.0288 & 1.158 & 1.65$\pm$0.07 & 9.34$\pm$0.20 \\
2007 Oct 24.0293 & 1.156 & 1.65$\pm$0.07 & 9.33$\pm$0.20 \\
2007 Oct 24.0366 & 1.139 & 1.73$\pm$0.07 & 9.29$\pm$0.20 \\
2007 Oct 24.0371 & 1.138 & 1.76$\pm$0.07 & 9.27$\pm$0.20 \\
2007 Oct 24.0443 & 1.123 & 1.83$\pm$0.07 & 9.23$\pm$0.20 \\
2007 Oct 24.0448 & 1.122 & 1.83$\pm$0.07 & 9.23$\pm$0.20 \\
2007 Oct 24.0520 & 1.109 & 1.95$\pm$0.07 & 9.16$\pm$0.20 \\
2007 Oct 24.0524 & 1.108 & 1.95$\pm$0.07 & 9.16$\pm$0.20 \\
2007 Oct 24.0597 & 1.097 & 2.07$\pm$0.07 & 9.10$\pm$0.20 \\
2007 Oct 24.0602 & 1.096 & 2.08$\pm$0.07 & 9.10$\pm$0.20 \\
2007 Oct 24.0696 & 1.084 & 2.27$\pm$0.07 & 9.00$\pm$0.20 \\
2007 Oct 24.0701 & 1.084 & 2.28$\pm$0.07 & 9.00$\pm$0.20 \\
2007 Oct 24.0773 & 1.076 & 2.46$\pm$0.07 & 8.91$\pm$0.20 \\
2007 Oct 24.0777 & 1.076 & 2.48$\pm$0.07 & 8.91$\pm$0.20 \\
2007 Oct 24.0850 & 1.069 & 2.69$\pm$0.07 & 8.82$\pm$0.20 \\
2007 Oct 24.0854 & 1.069 & 2.71$\pm$0.07 & 8.81$\pm$0.20 \\
2007 Oct 24.0927 & 1.065 & 3.00$\pm$0.07 & 8.69$\pm$0.20 \\
2007 Oct 24.0931 & 1.064 & 2.97$\pm$0.07 & 8.71$\pm$0.20 \\
2007 Oct 24.1005 & 1.061 & 3.28$\pm$0.07 & 8.60$\pm$0.20 \\
2007 Oct 24.1010 & 1.061 & 3.29$\pm$0.07 & 8.60$\pm$0.20 \\
2007 Oct 24.1093 & 1.059 & 3.84$\pm$0.07 & 8.43$\pm$0.20 \\
2007 Oct 24.1098 & 1.059 & 3.89$\pm$0.07 & 8.42$\pm$0.20 \\
2007 Oct 24.1160 & 1.059 & 4.21$\pm$0.07 & 8.33$\pm$0.20 \\
2007 Oct 24.1164 & 1.059 & 4.22$\pm$0.07 & 8.33$\pm$0.20 \\
2007 Oct 24.1225 & 1.060 & 4.62$\pm$0.07 & 8.23$\pm$0.20 \\
2007 Oct 24.1230 & 1.060 & 4.68$\pm$0.07 & 8.20$\pm$0.20 \\
2007 Oct 24.1292 & 1.063 & 5.21$\pm$0.07 & 8.10$\pm$0.20 \\
2007 Oct 24.1296 & 1.063 & 5.26$\pm$0.07 & 8.10$\pm$0.20 \\
2007 Oct 24.1359 & 1.066 & 5.88$\pm$0.07 & 7.97$\pm$0.20 \\
2007 Oct 24.1364 & 1.066 & 6.01$\pm$0.07 & 7.93$\pm$0.20 \\
2007 Oct 24.1426 & 1.071 & 6.60$\pm$0.07 & 7.80$\pm$0.20 \\
2007 Oct 24.1430 & 1.071 & 6.65$\pm$0.07 & 7.78$\pm$0.20 \\
2007 Oct 24.1492 & 1.076 & 7.43$\pm$0.07 & 7.62$\pm$0.20 \\
2007 Oct 24.1496 & 1.077 & 7.38$\pm$0.07 & 7.64$\pm$0.20 \\
\hline
\end{tabular}
\end{minipage}
\end{table}

\begin{table}
\begin{minipage}[t]{\columnwidth}
\caption{Selected Photometry from Other Observers}
\label{otherphotometry}
\centering
\renewcommand{\footnoterule}{}
\begin{tabular}{lccl}
\hline\hline
  UT Date
   & Mag.
   & Obs.\footnote{Observatory: 
      J47: Observatorio Nazaret (G. Muler), 0.20-m + CCD;
      J51: Observatorio Atlante, Tenerife (J. A. Henr\'iquez), 0.2-m + CCD;
      SPMN: Spanish Meteor and Fireball Network
      }
   & Reference \\
\hline
2007 Oct 24.06682 & 8.4 & J51  & \citet{spo07} \\
2007 Oct 24.06708 & 8.5 & J47  & \citet{spo07} \\
2007 Oct 24.06750 & 8.6 & J47  & \citet{spo07} \\
2007 Oct 24.06791 & 8.6 & J47  & \citet{spo07} \\
2007 Oct 24.07    & 8.4 & SPMN & \citet{tri08} \\
2007 Oct 24.07282 & 8.4 & J51  & \citet{spo07} \\
2007 Oct 24.07956 & 8.2 & J51  & \citet{spo07} \\
2007 Oct 24.09823 & 8.1 & J47  & \citet{spo07} \\
2007 Oct 24.09864 & 8.2 & J47  & \citet{spo07} \\
2007 Oct 24.09943 & 8.2 & J47  & \citet{spo07} \\
2007 Oct 24.10007 & 8.0 & J51  & \citet{spo07} \\
2007 Oct 24.10994 & 7.8 & J51  & \citet{spo07} \\
2007 Oct 24.11126 & 8.1 & J47  & \citet{spo07} \\
2007 Oct 24.11166 & 8.1 & J47  & \citet{spo07} \\
2007 Oct 24.11205 & 8.1 & J47  & \citet{spo07} \\
2007 Oct 24.11735 & 7.7 & J51  & \citet{spo07} \\
2007 Oct 24.12    & 7.8 & SPMN & \citet{tri08} \\
2007 Oct 24.12398 & 7.5 & J51  & \citet{spo07} \\
2007 Oct 24.12680 & 8.0 & J47  & \citet{spo07} \\
2007 Oct 24.12719 & 8.0 & J47  & \citet{spo07} \\
2007 Oct 24.13131 & 7.4 & J51  & \citet{spo07} \\
2007 Oct 24.13619 & 7.3 & J51  & \citet{spo07} \\
2007 Oct 24.15    & 7.4 & SPMN & \citet{tri08} \\
2007 Oct 24.18    & 7.0 & SPMN & \citet{tri08} \\
2007 Oct 24.20    & 6.7 & SPMN & \citet{tri08} \\
2007 Oct 24.21    & 6.5 & SPMN & \citet{tri08} \\
2007 Oct 24.22    & 6.3 & SPMN & \citet{tri08} \\
2007 Oct 24.23    & 6.0 & SPMN & \citet{tri08} \\
\hline
\end{tabular}
\end{minipage}
\end{table}

\begin{table}
\begin{minipage}[t]{\columnwidth}
\caption{Best-Fit Linear Parameters for Normalised Surface Brightness Profiles}
\label{lsbparams}
\centering
\renewcommand{\footnoterule}{}
\begin{tabular}{lrccrrcccc}
\hline\hline
 & \multicolumn{1}{c}{Sky\footnote{Average surface brightness of background sky in
     ADU~arcsec$^{-2}$}} 
 & \multicolumn{2}{c}{Region 1\footnote{Best-fit parameters for Region
     1 of surface brightness profiles in Figure~\ref{lsbprofiles} using
     $y=mx+b$ in log-log space, where parameters for Oct 31-Nov 3 are not calculated
     due to extensive inner coma saturation on these dates, and estimated uncertainties
     are dominated by variations in measured coma morphology from image to image during individual nights
     due to small sky brightness changes and field star interference}}
 & \multicolumn{2}{c}{Region 2\footnote{Best-fit parameters for Region 2, where estimated uncertainties
 	 are dominated by coma morphology variations during individual nights}}
 & \multicolumn{2}{c}{Region 3\footnote{Best-fit parameters for Region 3, where
     parameters for Nov 14-Dec 3 are not calculated due to the absence of
     this profile region on these dates, and estimated uncertainties
     are dominated by coma morphology variations during individual nights}}
 & \multicolumn{2}{c}{Region 4\footnote{Best-fit parameters for Region 4, where $m=0.0$ is assumed, and estimated uncertainties
     are dominated by coma morphology variations during individual nights}} \\
UT Date & \multicolumn{1}{c}{Level}
 & $m$ & $b$
 & \multicolumn{1}{c}{$m$} & \multicolumn{1}{c}{$b$}
 & $m$ & $b$
 & $m$ & $b$ \\
\hline
2007 Nov 01.08978 &  $2.1\pm0.1$ &   ---        &  ---        &  $-6.5\pm0.2$ & $15.5\pm0.2$ & $-0.9\pm0.1$ & $1.0\pm0.1$ & (0.0) & $-1.7\pm0.1$ \\
2007 Nov 02.17089 &  $1.6\pm0.1$ &   ---        &  ---        &  $-7.1\pm0.2$ & $17.1\pm0.2$ & $-0.8\pm0.1$ & $0.7\pm0.1$ & (0.0) & $-1.8\pm0.1$ \\
2007 Nov 03.14726 &  $1.4\pm0.1$ &   ---        &  ---        &  $-7.1\pm0.2$ & $17.3\pm0.2$ & $-0.8\pm0.1$ & $0.5\pm0.1$ & (0.0) & $-1.9\pm0.1$ \\
2007 Nov 03.97323 &  $0.8\pm0.1$ & $-0.9\pm0.1$ & $1.8\pm0.1$ &  $-9.1\pm0.2$ & $22.7\pm0.2$ & $-0.8\pm0.1$ & $0.4\pm0.1$ & (0.0) & $-2.1\pm0.1$ \\
2007 Nov 05.13939 &  $0.9\pm0.1$ & $-0.9\pm0.1$ & $1.6\pm0.1$ &  $-7.7\pm0.2$ & $19.2\pm0.2$ & $-0.7\pm0.1$ & $0.0\pm0.1$ & (0.0) & $-2.1\pm0.1$ \\
2007 Nov 06.05627 &  $0.8\pm0.1$ & $-0.8\pm0.1$ & $1.4\pm0.1$ &  $-7.2\pm0.2$ & $18.1\pm0.2$ & $-0.7\pm0.1$ & $0.2\pm0.1$ & (0.0) & $-2.1\pm0.1$ \\
2007 Nov 14.02572 &  $1.1\pm0.1$ & $-0.7\pm0.1$ & $1.4\pm0.1$ &  $-8.6\pm0.2$ & $23.5\pm0.2$ &   ---        &  ---        & (0.0) & $-1.6\pm0.1$ \\
2007 Nov 15.12698 &  $1.0\pm0.1$ & $-0.7\pm0.1$ & $1.4\pm0.1$ & $-10.6\pm0.2$ & $29.3\pm0.2$ &   ---        &  ---        & (0.0) & $-1.6\pm0.1$ \\
2007 Nov 16.21182 &  $0.9\pm0.1$ & $-0.7\pm0.1$ & $1.4\pm0.1$ &  $-9.4\pm0.2$ & $26.2\pm0.2$ &   ---        &  ---        & (0.0) & $-1.6\pm0.1$ \\
2007 Nov 16.96157 &  $1.1\pm0.1$ & $-0.7\pm0.1$ & $1.3\pm0.1$ &  $-8.3\pm0.2$ & $23.2\pm0.2$ &   ---        &  ---        & (0.0) & $-1.5\pm0.1$ \\
2007 Nov 23.13081 & $12.8\pm0.1$ & $-0.3\pm0.1$ & $0.4\pm0.1$ &  $-1.2\pm0.2$ & $~3.2\pm0.2$ &   ---        &  ---        & (0.0) & $-0.4\pm0.1$ \\
2007 Dec 02.12217 &  $1.2\pm0.1$ & $-0.6\pm0.1$ & $1.2\pm0.1$ &  $-4.7\pm0.2$ & $13.6\pm0.2$ &   ---        &  ---        & (0.0) & $-0.9\pm0.1$ \\
2007 Dec 03.02399 &  $0.7\pm0.1$ & $-0.6\pm0.1$ & $1.3\pm0.1$ &  $-6.9\pm0.2$ & $20.3\pm0.2$ &   ---        &  ---        & (0.0) & $-1.1\pm0.1$ \\
\hline
\end{tabular}
\end{minipage}
\end{table}

\begin{table}
\begin{minipage}[t]{\columnwidth}
\caption{Coma Size Measurements}
\label{expansion}
\centering
\renewcommand{\footnoterule}{}
\begin{tabular}{lcccccc}
\hline\hline
 & & & R1-R2 & R2-R3 & R3-R4 & R2-R4 \\
UT Date & JD-2450000.5
 & km/arcsec\footnote{Conversion factor used to compute physical distances at geocentric
           distance of the comet from angular distances}
 & Distance\footnote{Distance from nucleus of intersection point between
                   Regions 1 and 2 of linear surface brightness profiles
		   in 10$^5$~km, where distances on Oct 31-Nov 3 are not calculated due to the unavailability
		   of Region 1 parameters due to extensive inner coma saturation on these dates}
 & Distance\footnote{Distance from nucleus of intersection point between
                   Regions 2 and 3
		   in 10$^5$~km}
 & Distance\footnote{Distance from nucleus of intersection point between
                   Regions 2 and 4x
		   in 10$^5$~km, where distances on Nov 14-Dec 3 are not calculated due to the
		   absence of Region 2 on these dates}
 & Distance\footnote{Distance from nucleus of intersection point between
                   Regions 3 and 4
		   in 10$^5$~km} \\
\hline
2007 Nov 01.08978 & 4405.08978 & 1178.1 &      ---      & $4.3\pm0.3$ & $12.3\pm0.3$ & ~~$5.0\pm0.3$ \\
2007 Nov 02.17089 & 4406.17089 & 1178.1 &      ---      & $4.9\pm0.3$ & $14.7\pm0.3$ & ~~$5.5\pm0.3$ \\
2007 Nov 03.14726 & 4407.14726 & 1178.1 &      ---      & $5.3\pm0.3$ & $16.2\pm0.3$ & ~~$6.0\pm0.3$ \\
2007 Nov 03.97323 & 4407.97323 & 1178.1 & ~~$4.2\pm0.3$ & $5.6\pm0.3$ & $17.3\pm0.3$ & ~~$6.2\pm0.3$ \\
2007 Nov 05.13939 & 4409.13939 & 1178.1 & ~~$4.6\pm0.3$ & $6.3\pm0.3$ & $19.3\pm0.3$ & ~~$6.9\pm0.3$ \\
2007 Nov 06.05627 & 4410.05627 & 1178.1 & ~~$4.8\pm0.3$ & $6.7\pm0.3$ & $18.7\pm0.3$ & ~~$7.1\pm0.3$ \\
2007 Nov 14.02572 & 4418.02572 & 1185.4 & ~~$7.8\pm0.3$ &     ---     &      ---     & ~~$9.8\pm0.3$ \\
2007 Nov 15.12698 & 4419.12698 & 1185.4 & ~~$8.2\pm0.3$ &     ---     &      ---     &  $10.0\pm0.3$ \\
2007 Nov 16.21182 & 4420.21182 & 1185.4 & ~~$8.5\pm0.3$ &     ---     &      ---     &  $10.5\pm0.3$ \\
2007 Nov 16.96157 & 4420.96157 & 1185.4 & ~~$8.6\pm0.3$ &     ---     &      ---     &  $10.8\pm0.3$ \\
2007 Nov 23.13081 & 4427.13081 & 1199.9 &  $10.2\pm0.3$ &     ---     &      ---     &  $12.4\pm0.3$ \\
2007 Dec 02.12217 & 4436.12217 & 1236.3 &  $13.1\pm0.3$ &     ---     &      ---     &  $16.1\pm0.3$ \\
2007 Dec 03.02399 & 4437.02399 & 1236.3 &  $13.6\pm0.3$ &     ---     &      ---     &  $16.4\pm0.3$ \\
\hline
\end{tabular}
\end{minipage}
\end{table}

\begin{table}
\begin{minipage}[t]{\columnwidth}
\caption{False Nucleus Distance Measurements}
\label{fragmotion}
\centering
\renewcommand{\footnoterule}{}
\begin{tabular}{lcrc}
\hline\hline
 & & \multicolumn{2}{c}{Sep. Dist.\footnote{Separation distance between true nucleus and false nucleus}} \\
UT Date & JD-2450000.5 & arcsec & $10^5$~km \\
\hline
2007 Nov 13.92576 & 4417.92576 & 186$\pm$7 & 2.2$\pm$0.1 \\
2007 Nov 14.02572 & 4418.02572 & 184$\pm$7 & 2.2$\pm$0.1 \\
2007 Nov 14.11186 & 4418.11186 & 189$\pm$7 & 2.2$\pm$0.1 \\
2007 Nov 14.21586 & 4418.21586 & 182$\pm$7 & 2.2$\pm$0.1 \\
2007 Nov 14.92619 & 4418.92619 & 200$\pm$7 & 2.4$\pm$0.1 \\
2007 Nov 15.01977 & 4419.01977 & 182$\pm$7 & 2.2$\pm$0.1 \\
2007 Nov 15.12698 & 4419.12698 & 199$\pm$7 & 2.4$\pm$0.1 \\
2007 Nov 15.21417 & 4419.21417 & 199$\pm$7 & 2.4$\pm$0.1 \\
2007 Nov 15.92746 & 4419.92746 & 201$\pm$7 & 2.4$\pm$0.1 \\
2007 Nov 16.02321 & 4420.02321 & 206$\pm$7 & 2.4$\pm$0.1 \\
2007 Nov 16.11663 & 4420.11663 & 199$\pm$7 & 2.4$\pm$0.1 \\
2007 Nov 16.21183 & 4420.21183 & 203$\pm$7 & 2.4$\pm$0.1 \\
2007 Nov 16.92535 & 4420.92535 & 214$\pm$7 & 2.5$\pm$0.1 \\
2007 Nov 16.96157 & 4420.96157 & 215$\pm$7 & 2.6$\pm$0.1 \\
2007 Nov 17.00595 & 4421.00595 & 204$\pm$7 & 2.4$\pm$0.1 \\
2007 Nov 17.03829 & 4421.03829 & 210$\pm$7 & 2.5$\pm$0.1 \\
2007 Nov 23.10448 & 4427.10448 & 255$\pm$7 & 3.1$\pm$0.1 \\
2007 Nov 23.13082 & 4427.13082 & 266$\pm$7 & 3.2$\pm$0.1 \\
2007 Nov 23.16376 & 4427.16376 & 260$\pm$7 & 3.1$\pm$0.1 \\
2007 Nov 23.19162 & 4427.19162 & 274$\pm$7 & 3.3$\pm$0.1 \\
2007 Dec 01.95611 & 4435.95611 & 333$\pm$7 & 4.1$\pm$0.1 \\
2007 Dec 02.12217 & 4436.12217 & 329$\pm$7 & 4.1$\pm$0.1 \\
2007 Dec 02.83004 & 4436.83004 & 338$\pm$7 & 4.2$\pm$0.1 \\
2007 Dec 03.02399 & 4437.02399 & 327$\pm$7 & 4.1$\pm$0.1 \\
\hline
\end{tabular}
\end{minipage}
\end{table}

\clearpage

\bsp

\label{lastpage}

\end{document}